\documentclass[aps,prd,twocolumn,subscriptaddress,amsmath,amssymb
]{revtex4-2}

\usepackage[compatibility=false]{caption}

\usepackage{orcidlink}
\usepackage{float}
\usepackage{graphicx}
\usepackage{dcolumn}
\usepackage{booktabs}
\usepackage{placeins}
\usepackage{style}
\usepackage{comment}
\usepackage{tabularx}

\usepackage{booktabs}
\usepackage{multirow}
\usepackage{slashed}

\usepackage{url}
\usepackage{hyperref}

\usepackage{ragged2e}

\DeclareCaptionJustification{myjustified}{\justifying}
\begin{document}

\title{Sneutrino Tribrid Inflation in Flipped $\mathbf{SU(5)}$: Confronting ACT DR6 and Planck Observations}

\author{Farishta Israr \orcidlink{0009-0007-1044-7630}}
\email{farishtaisrar@yahoo.com}
\affiliation{Department of Physics, Quaid-i-Azam University, Islamabad, 45320, Pakistan}

\author{Mansoor Ur Rehman
\orcidlink{0000-0002-1780-1571}}
\email{m.rehman@iu.edu.sa}

\author{Saleh O. Allehabi
\orcidlink{0000-0002-3792-8561}}
\email{s.allehabi@iu.edu.sa}
\affiliation{Department of Physics, Faculty of Science, Islamic University of Madinah, 42351 Madinah, Saudi Arabia}

\begin{abstract}
We construct a realization of sneutrino tribrid inflation within the
$R$-symmetric flipped $SU(5)$ grand unified theory. A vector-like pair of
matter multiplets, $10_V+\overline{10}_V$, provides a $D$-flat inflaton
direction along the right-handed sneutrino component, with an additional
$Z_2$ symmetry forbidding a direct inflaton mass term so that the leading
inflationary interaction is a non-renormalizable K\"ahler-driven operator.
Unlike previous studies, we treat the superpotential cutoff scale $M_s$ as
independent of the reduced Planck mass, $m_P$, substantially enlarging the viable
parameter space. Confronting the model with the ACT DR6/Planck
determination $n_s=0.9734\pm0.0034$ and the $2\sigma$ bound on the running
of the spectral index, we identify broad regions of viable parameter space
in which the flipped $SU(5)$ symmetry-breaking scale can approach the
conventional GUT scale, $M\simeq M_{\rm GUT}$, and the tensor-to-scalar
ratio can reach observable values, $r\gtrsim10^{-3}$, within reach of
forthcoming CMB $B$-mode experiment such as LiteBIRD. We
further study the post-inflationary dynamics, identifying the inflaton
with the lightest right-handed sneutrino in a conventional type-I
seesaw sector, whose out-of-equilibrium decay generates a lepton asymmetry that is
converted into the observed baryon asymmetry via non-thermal leptogenesis
and electroweak sphalerons, for a reheating temperature consistent with
both successful leptogenesis and the gravitino constraint.

\end{abstract}

\maketitle

\section{Introduction}

Cosmic inflation provides a compelling explanation for the flatness, horizon, and monopole problems of the standard Big Bang cosmology while simultaneously generating the nearly scale-invariant spectrum of primordial density perturbations observed in the cosmic microwave background (CMB)~\cite{Guth:1980zm,Linde:1981mu,Albrecht:1982wi,Mukhanov:1981xt,Hawking:1982cz,Starobinsky:1982ee,Guth:1982ec}. Embedding inflation within a supersymmetric grand unified theory (SUSY GUT) offers an attractive framework for connecting the dynamics of the early Universe with high-energy particle physics, thereby providing a unified setting to address gauge unification, neutrino masses, and the origin of the baryon asymmetry.

Among the various realizations, supersymmetric hybrid inflation~\cite{Dvali:1994ms,Copeland:1994vg,Senoguz:2003zw,Senoguz:2004vu} provides one of the most attractive and economical frameworks.  In this scenario, inflation is driven by the $F$-term vacuum energy associated with a gauge-singlet field and terminates through a waterfall phase transition that spontaneously breaks the underlying gauge symmetry. In the global supersymmetric limit, the tree-level inflaton potential is exactly flat, with the required slope generated radiatively. This minimal scenario typically predicts a scalar spectral index in the range $n_s\simeq0.98$--$1.00$, somewhat larger than the value favored by current CMB observations. Moreover, the inclusion of supergravity (SUGRA) corrections arising from the minimal canonical K\"ahler potential generally exacerbates this discrepancy~\cite{Linde:1997sj,Senoguz:2004vu}. Several mechanisms have been proposed to reconcile the model with observations, including soft supersymmetry-breaking terms~\cite{Rehman:2009nq,Rehman:2025fja} and higher-dimensional operators in a non-minimal K\"ahler potential~\cite{Bastero-Gil:2006zpr,urRehman:2006hu}. 
More recently, the ACT DR6 observations, when combined with Planck data, favor a scalar spectral index of $n_s=0.9734\pm0.0034$~\cite{ACTDR6:2025}, noticeably higher than the Planck-only determination, $n_s=0.9649\pm0.0042$~\cite{Planck:2018vyg,Planck:2018jri}, while maintaining a stringent upper bound on the tensor-to-scalar ratio. This shift has renewed interest in supersymmetric hybrid inflation and its variants, motivating a re-examination of their predictions within realistic GUT frameworks.
Recent studies in this direction include Refs.~\cite{Ahmed:2025eip,Ahmed:2026gqq,Ijaz:2026ear,Pallis:2026qci,Pallis:2026cyz,Pallis:2025vxo,Pallis:2025gii,Pallis:2025nrv}.

In tribrid inflation~\cite{Antusch:2004hd,Antusch:2009vg,Antusch:2009ef,Antusch:2012bp,Antusch:2012jc}, the roles of the three scalar fields are clearly separated: a gauge-singlet field supplies the vacuum energy, matter fields serve as the inflaton, and Higgs fields are responsible for terminating inflation through a waterfall transition. This framework is particularly appealing in grand unified theories, as it allows gauge non-singlet matter multiplets to play an active role in the inflationary dynamics while simultaneously linking cosmology with particle physics~\cite{Antusch:2010va,Masoud:2021prr,Moursy:2020sit,Ahmed:2025crx}. In particular, identifying the inflaton with a right-handed sneutrino establishes an intriguing connection between inflation, the seesaw mechanism, and the origin of the observed baryon asymmetry through leptogenesis~\cite{Antusch:2010mv}.

To realize successful inflation, the inflaton trajectory must satisfy the $D$-flatness condition in order to eliminate the large quartic contributions arising from the gauge $D$-term potential. While this condition is automatically satisfied when the inflaton is a gauge singlet, as in the sneutrino inflation model based on conventional $SU(5)$~\cite{Masoud:2019gxx}, it requires a nontrivial construction for a gauge non-singlet inflaton. A simple and elegant solution is to introduce a vector-like matter multiplet, whose $D$-flat combination with the matter fields provides a suitable inflationary direction. This mechanism has been successfully implemented in Pati--Salam and $SO(10)$ grand unified theories~\cite{Antusch:2010va}, in the $U(1)_{B-L}$ extension of the MSSM~\cite{Masoud:2021prr,Moursy:2020sit}, and more recently in the left--right symmetric model~\cite{Ahmed:2025crx}.

In this paper, we investigate the realization of tribrid inflation within the flipped $SU(5)$ grand unified theory \cite{Barr:1981qv,Derendinger:1983aj} based on the gauge symmetry $SU(5) \times U(1)_X$. Unlike conventional $SU(5)$, the flipped $SU(5)$ framework naturally resolves the doublet--triplet splitting problem through the missing partner mechanism~\cite{Antoniadis:1987dx}, thereby eliminating the need for fine-tuning. Since the electroweak Higgs doublets remain massless, the MSSM $\mu$-term can be generated via the Giudice--Masiero mechanism~\cite{Giudice:1988yz}. Moreover, the underlying $R$-symmetry enhances the theoretical consistency of the model by forbidding the dangerous dimension-four and dimension-five baryon-number violating operators that would otherwise induce rapid proton decay~\cite{Mehmood:2020irm}.
The flipped $SU(5)$ model also enjoys several distinctive theoretical advantages. In particular, unlike the $U(1)_{B-L}$, conventional $SU(5)$, and $SO(10)$ frameworks, it is free from the cosmological problems associated with the formation of topological defects, including $B\!-\!L$ cosmic strings and magnetic monopoles. Moreover, flipped $SU(5)$ admits a natural realization within $F$-theory, establishing an intriguing link between grand unification and string theory~\cite{Jiang:2008xrg,Jiang:2009za,Li:2009qlh,Chen:2010tp,Chung:2010bn,Antoniadis:2025yau}.

To realize a $D$-flat inflaton direction, we extend the flipped $SU(5)$ matter sector by introducing a vector-like pair of multiplets, $10_V+\overline{10}_V$, in addition to the ordinary matter multiplets $10$. The inflaton is identified with a $D$-flat combination of these fields, chosen to lie along the right-handed sneutrino direction. An additional $Z_2$ symmetry is imposed to forbid a direct inflaton mass term. As a consequence, radiative corrections to the inflationary potential are highly suppressed, placing our construction within the class of K\"ahler-driven tribrid inflation models~\cite{Antusch:2012jc}, where the inflationary dynamics are dominated by higher-dimensional supergravity corrections originating from the K\"ahler potential. In contrast to the conventional framework, which assumes the reduced Planck mass $m_P$ as the unique cutoff scale, we distinguish between the cutoff scale $M_s$ entering the superpotential and the reduced Planck mass governing the supergravity corrections. This generalized setup substantially enlarges the viable parameter space compared to previous studies. 
We perform a comprehensive scan over this extended parameter space, confronting the model with the latest ACT DR6 and Planck CMB observations while simultaneously imposing theoretically motivated consistency conditions. This analysis identifies broad regions of viable parameter space that satisfy all current observational constraints. Following inflation, the sneutrino inflaton reheats the Universe through its decay, naturally generating the observed baryon asymmetry via non-thermal leptogenesis.

The remainder of this paper is organized as follows. In
Sec.~\ref{sec:RSU5}, we present the field content, symmetry
assignments, and superpotential of the $R$-symmetric flipped $SU(5)$ model,
including the vector-like matter multiplets required for a $D$-flat
inflaton direction. Sec.~\ref{sec:global_potential} derives the global
supersymmetric $F$- and $D$-term scalar potentials and identifies the
$D$-flat sneutrino inflationary trajectory. In
Sec.~\ref{sec:SUGRA_potential}, we incorporate supergravity corrections
through a non-minimal K\"ahler potential, obtain the effective single-field
inflaton potential, and show that radiative and soft supersymmetry-breaking
corrections remain subdominant. Sec.~\ref{sec:inflation-dynamics} develops
the slow-roll formalism and derives analytic expressions for the
inflationary observables. In Sec.~\ref{sec:numerical_results},
we perform a comprehensive numerical scan of the parameter space and
confront the model predictions with the latest ACT DR6 and Planck
constraints. Sec.~\ref{sec:reheating} discusses inflaton decay, reheating,
and the generation of the baryon asymmetry through non-thermal
leptogenesis. Finally, we summarize our results and conclude in
Sec.~\ref{sec:conclusions}.

\section{$R$-symmetric flipped $SU(5)$ model} \label{sec:RSU5}
In this section, we present the supersymmetric flipped $SU(5)$ model that realizes sneutrino tribrid inflation. We first summarize the field content and symmetry assignments and then construct the superpotential consistent with the underlying gauge and discrete symmetries. 

The three generations of matter superfields are embedded into the flipped
$SU(5)\times U(1)_X$ representations
$10_i$, $\overline5_i$, and $1_i$,
which contain the Standard Model fermions together with the right-handed
neutrinos.
The GUT symmetry is broken by the Higgs multiplets
$(10_H,\overline{10}_H)$,
while electroweak symmetry breaking is achieved by the Higgs pair
$(5_h,\overline5_h)$.
To realize sneutrino tribrid inflation, we further introduce a vector-like
pair of matter multiplets,
$10_V+\overline{10}_V$.
Their presence allows the construction of a $D$-flat inflaton direction,
identified with a linear combination of the right-handed sneutrino
components contained in $10_i$ and $\overline{10}_V$.
Throughout inflation, the GUT Higgs fields
$10_H$ and $\overline{10}_H$
are stabilized at the origin and subsequently act as the waterfall fields.

Besides the gauge symmetry, we impose
matter parity $\mathbb Z_2$
to forbid the dangerous baryon- and lepton-number violating operators
and to ensure the stability of the lightest supersymmetric particle.

Since the vector-like multiplet $10_V$ carries the same gauge quantum
numbers as the ordinary matter multiplets $10_i$,
the gauge symmetry alone would allow the renormalizable mixing
$M_i10_i\overline{10}_V$.
Such a term would lift the desired $D$-flat inflaton direction and
generate an inflaton mass of order the GUT scale,
thereby preventing slow-roll inflation.
To eliminate these unwanted mixings,
we impose an additional $Z_2$ symmetry under which
$10_V$ and $\overline{10}_V$
transform differently from the ordinary matter multiplets.

An immediate consequence of the additional $Z_2$ symmetry is that the
renormalizable hybrid-inflation coupling
$S10_H\overline{10}_H$
is forbidden. The leading allowed interaction in the inflationary sector is
therefore the non-renormalizable operator,
$\frac{S(10_H\overline{10}_H)^2}{M_s^2}$,
which naturally leads to a tribrid inflationary structure. An alternative
construction that retains the renormalizable coupling
$S10_H\overline{10}_H$
was considered in Ref.~\cite{Antusch:2010va}. After GUT symmetry breaking,
the higher-dimensional interactions involving
$10_a$, $\overline{10}_V$, $10_H$, and $\overline{10}_H$
also generate masses for the vector-like multiplets.

\begin{table*}[t]
\caption{\label{sf}
Superfield content of the flipped $SU(5)$ model, together with the
$SU(5)\times U(1)_X$ representations, decomposition under the Standard Model
gauge group, and charge assignments under matter parity $\mathbb{Z}_2$,
the additional $Z_2$ symmetry, and $U(1)_R$.
}
\begin{ruledtabular}
\begin{tabular}{cclccc}
\textrm{Representation}&\textrm{  $SU(5)\times{U(1)_X}$}&
\textrm{$SU(3)_c\times SU(2)_L \times U(1)_Y$} &$\mathbb{Z}_2$ &$Z_2$ & $U(1)_R$  \\
\hline
 $10_i$&$  (10,  1)$  & $Q_i\  ( 3, 2, 1/6)$
 + $D^c_i\  ( \overline{3}, 1, 1/3)$  
 + $N^c_i\  ( 1, 1, 0)$&$-1$ &$+1$ & $0$  \\
 \hline
$\bar{5}_i\ $&$ (\bar{5},  -3)$ & $U^c_i\  ( \overline{3}, 1, -2/3)$
+ $L_i \ ( 1, 2, -1/2)$&$-1$&$+1$&$0$ \\
\hline
$1_i\ $&$ (1,  5)$ & $E^c_i \ ( 1, 1, 1)$&$-1$&$+1$& $0$  \\
\hline
$5_{h}\ $& $ (5, -2)$  & $H_{d}\ (1,2,-1/2)$ 
 + $H_{T}\ (3,1,-1/3)$&$+1$ &$+1$&$1$ \\
 \hline 
$\bar{5}_{h}\ $&$ (\bar{5}, 2)$  & $ {H}_{u}\ (1,2,+1/2)$  
 + $\bar{H}_{T}\ (\bar{3},1,+1/3)$&$+1$&$+1$& $1$ \\
 \hline 
 $10_H\ $&$ (10,  1)$  & $Q_H \ ( 3, 2, 1/6)$
 + $D^c_H \ ( \overline{3}, 1, 1/3)$  
 + $N^c_H\  ( 1, 1, 0)$&$+1$&$-1$ &$0$ \\
  \hline 
 $\overline{10}_H\ $&$ (\overline{10},  -1)$  & $\overline{Q}_H \ ( \overline{3}, 2, -1/6)$
 + $\overline{D}^c_H \ ( 3, 1, -1/3)$  
 + $\overline{N}^c_H\  ( 1, 1, 0)$&$+1$ &$+1$&$0$ \\
 \hline
 $S\ $& $(1,  0)$  & $S\ (1,1,0)$&$+1$&$+1$ &$1$ \\
  \hline
 $10_V\ $& $(10,  1)$  &  $Q_V \ ( 3, 2, 1/6)$
 + $D^c_V \ ( \overline{3}, 1, 1/3)$  
 + $N^c_V\  ( 1, 1, 0)$  &$-1$&$+1$& $0$ \\
  \hline
 $\overline{10}_V\ $& $(\overline{10},  -1)$  &  $\overline{Q}_V \ ( \overline{3}, 2, -1/6)$
 + $\overline{D}^c_V \ ( 3, 1, -1/3)$  
 + $\overline{N}^c_V\  ( 1, 1, 0)$  &$-1$&$-1$&$1$ \\
\end{tabular}
\end{ruledtabular}
\end{table*}
The complete superfield content together with the corresponding gauge,
discrete, and $R$-charge assignments is summarized in
Table~\ref{sf}.
The superpotential of the model under these symmetries can be written as
\begin{align}\label{W}
W &\supset
S \left( \frac{(10_H \, \overline{10}_H)^2}{M_s^2} - \mu_s^2 \right)\nonumber \\
&+ \lambda_1 \, 10_H \, 10_H \, 5_h 
+ \lambda_2 \, \overline{10}_H \, \overline{10}_H \, \overline{5}_h \nonumber \\
&+ y_{ab}^{(d)} \, 10_a \, 10_b \, 5_h 
+ y_{aj}^{(u,\nu)} \, 10_a \, \overline{5}_j \, \overline{5}_h
+ y_{ij}^{(e)} \, 1_i \, \overline{5}_j \, 5_h \nonumber \\
& + \frac{\gamma_a}{M_s} \, (10_a \, \overline{10}_V)\,(10_H \, \overline{10}_H) 
+ \frac{\beta_a}{M_s}  (10_a \, \overline{10}_H)( 10_H\, \overline{10}_V),
\end{align}
with
\begin{align}
10_a & = (10_i, 10_V),\, y_{ab}^{(d)} = (y_{ij}^{(d)}, y_{iV}, \, y_V), \\ y_{aj}^{(u,\nu)} & = ( y_{ij}^{(u,\nu)}, \bar{y}_{Vj}), \, a,\,b=(i,V),(j,V), \, (i,j) = 1,2,3.
\end{align}
Here,
$\lambda_1$, $\lambda_2$,
$\gamma_a$,
and $\beta_a$
are dimensionless couplings,
while
$y^{(d)}$,
$y^{(u,\nu)}$,
and
$y^{(e)}$
denote the Yukawa matrices.
The parameters
$\mu_s$
and
$M_s$
represent the inflationary mass scale and the effective cutoff scale,
respectively.

The first term in Eq.~(\ref{W}) determines the supersymmetric vacuum in which
the Standard Model singlet components of $10_H$ and $\overline{10}_H$
acquire GUT-scale vacuum expectation values, thereby breaking
$SU(5)\times U(1)_X$ to the Standard Model gauge group. The second line
contains the Higgs interactions responsible for realizing the missing partner
mechanism. Compared with conventional $SU(5)$, this mechanism naturally
resolves the doublet--triplet splitting problem~\cite{Antoniadis:1987dx}.
Once $N_H^c$ and $\overline{N}_H^c$ acquire their GUT-scale VEVs, the
color-triplet fields $(H_T,\overline H_T)$ contained in
$(5_h,\overline5_h)$ pair with $(D_H^c,\overline D_H^c)$ from
$(10_H,\overline{10}_H)$ and obtain GUT-scale masses through the couplings
$\lambda_1$ and $\lambda_2$. In contrast, the electroweak Higgs doublets
remain massless at this stage. The MSSM $\mu$-term may subsequently be
generated through the Giudice--Masiero mechanism~\cite{Giudice:1988yz}.
The absence of light color-triplet partners strongly suppresses the dangerous
dimension-five proton decay operators characteristic of conventional
supersymmetric $SU(5)$, although potentially observable proton decay may still
arise in realistic flipped $SU(5)$ constructions~\cite{Mehmood:2020irm}.

The Yukawa interactions in the third line of Eq.~(\ref{W}) generate the
quark and lepton masses after electroweak symmetry breaking. The first and
last two terms constitute the tribrid inflationary sector. Unlike conventional
hybrid inflation, the scalar component of the gauge-singlet superfield $S$
is stabilized at the origin, while its nonvanishing $F$-term provides the
vacuum energy that drives inflation. The inflaton is identified with the
$D$-flat right-handed sneutrino direction formed by the singlet components
of $10_i$ and $\overline{10}_V$. The GUT Higgs multiplets $10_H$ and
$\overline{10}_H$ remain stabilized at the origin during inflation and act
as the waterfall fields. When the inflaton reaches its critical value, the
waterfall direction becomes tachyonic, terminating inflation and triggering
the breaking of $SU(5)\times U(1)_X$ to the Standard Model gauge group.

The resulting construction belongs to the class of K\"ahler-driven tribrid
inflation models~\cite{Antusch:2012jc}, in which the slope of the inflaton
potential is generated predominantly by supergravity corrections induced by
higher-dimensional operators in the K\"ahler potential, while radiative
corrections remain subdominant. In contrast to the model-independent analyses
of Refs.~\cite{Antusch:2004hd,Antusch:2012jc}, where the cutoff scale is
identified with the reduced Planck mass, we allow the superpotential cutoff
$M_s$ to differ from the reduced Planck mass $m_P$ governing the supergravity
expansion. This distinction permits a substantially broader parameter space
to be explored and provides a realistic realization of gauge non-singlet
sneutrino tribrid inflation within flipped $SU(5)$.
\section{Global supersymmetric scalar potential}
\label{sec:global_potential}

In this section, we derive the global supersymmetric scalar potential associated
with the inflationary sector of the model. 

\subsection{Inflationary superpotential and supersymmetric vacuum}

The superpotential in Eq.~(\ref{W}) contains the interactions responsible for
GUT symmetry breaking, fermion masses, and inflation. Retaining only the terms
relevant for the inflationary dynamics, we write
\begin{align}
W_{\rm inf} ={}&
S\left[
\frac{\left(10_H\cdot\overline{10}_H\right)^2}{M_s^2}
-\mu_s^2
\right]
\nonumber\\
&+
\frac{\gamma_a}{M_s}
\left(10_a\cdot\overline{10}_V\right)
\left(10_H\cdot\overline{10}_H\right)
\nonumber\\
&+
\frac{\beta_a}{M_s}
\left(10_a\cdot\overline{10}_H\right)
\left(\overline{10}_V\cdot10_H\right),
\label{Winf}
\end{align}
where summation over the index $a$ is understood.

The supersymmetric vacuum is determined by the simultaneous vanishing of all
$F$- and $D$-terms. Using the $SU(5)\times U(1)_X$ gauge freedom, the vacuum
expectation values of the GUT Higgs multiplets may be aligned along their
Standard-Model-singlet components. The vacuum configuration is then given by
\begin{equation}
\begin{aligned}
& \langle S \rangle = 0, \quad
\langle 10_a \rangle = \langle \overline{10}_V \rangle = 0, \\
& \langle 10_H \cdot \overline{10}_H \rangle
= \langle {N}_H^c \,\overline{N}_H^c \rangle
= M^2 ,
\end{aligned}
\label{eq:vevs_global}
\end{equation}

The condition $F_S=0$ determines the symmetry-breaking scale,
\begin{equation}
M \equiv \sqrt{\mu_s M_s}.
\label{eq:Mscale}
\end{equation}
At this vacuum, supersymmetry is restored, while the nonzero expectation values of $N_H^c$ and $\overline{N}_H^c$ spontaneously break $SU(5)\times U(1)_X$ to the Standard Model gauge group.

\subsection{F-term scalar potential}

For a canonical K\"ahler potential, the global supersymmetric
$F$-term scalar potential is
\begin{equation}
V_F =
|F_S|^2
+ \left\|F_{10_H}\right\|^2
+ \left\|F_{\overline{10}_H}\right\|^2
+ \sum_a \left\|F_{10_a}\right\|^2
+ \left\|F_{10_V}\right\|^2 ,
\label{eq:Fpotential}
\end{equation}
where
\begin{equation}
F_Z^* = -\frac{\partial W_{\mathrm{inf}}}{\partial Z}.
\label{eq:Fterm}
\end{equation}

The explicit expression following from Eq.~(\ref{Winf}) is
\begin{align}
V_F ={}&
\left|
\frac{(10_H\!\cdot\!\overline{10}_H)^2}{M_s^2}
-\mu_s^2
\right|^2
\nonumber\\[1mm]
&+
\left\|
\frac{2S}{M_s^2}
(10_H\!\cdot\!\overline{10}_H)\,
\overline{10}_H
+
\frac{\gamma_a}{M_s}
(10_a\!\cdot\!\overline{10}_V)\,
\overline{10}_H
\right.
\nonumber\\[-1mm]
&\hspace{1.6cm}\left.
+
\frac{\beta_a}{M_s}
(10_a\!\cdot\!\overline{10}_H)\,
\overline{10}_V
\right\|^2
\nonumber \\[1mm]
&+
\left\|
\frac{2S}{M_s^2}
(10_H\!\cdot\!\overline{10}_H)\,
10_H
+
\frac{\gamma_a}{M_s}
(10_a\!\cdot\!\overline{10}_V)\,
10_H
\right.
\nonumber  \\[-1mm]
&\hspace{1.6cm}\left.
+
\frac{\beta_a}{M_s}
(\overline{10}_V\!\cdot\!10_H)\,
10_a
\right\|^2
\nonumber \\[1mm]
&+
\sum_a
\left\|
\frac{\gamma_a}{M_s}
(10_H\!\cdot\!\overline{10}_H)\,
\overline{10}_V
+
\frac{\beta_a}{M_s}
(\overline{10}_V\!\cdot\!10_H)\,
\overline{10}_H
\right\|^2
\nonumber \displaybreak[3]  \\[1mm]
&+
\left\|
\sum_a
\left[
\frac{\gamma_a}{M_s}
(10_H\!\cdot\!\overline{10}_H)\,
10_a
+
\frac{\beta_a}{M_s}
(10_a\!\cdot\!\overline{10}_H)\,
10_H
\right]
\right\|^2 .
\label{eq:VF}
\end{align}
The $SU(5)$-invariant contraction between a $10$-plet and an
$\overline{10}$-plet is defined by
\begin{equation}
10\cdot\overline{10}
\equiv
\frac{1}{2}
10^{\alpha\beta}\overline{10}_{\alpha\beta},
\label{eq:10_contraction}
\end{equation}
where
\begin{equation}
10^{\alpha\beta}=-10^{\beta\alpha},
\qquad
\overline{10}_{\alpha\beta}
=
-\overline{10}_{\beta\alpha},
\end{equation}
and $\alpha,\beta=1,\ldots,5$ are $SU(5)$ indices. The corresponding norm is
defined as
\begin{equation}
\|10\|^2
\equiv
\frac{1}{2}
10^{\alpha\beta\ast}10^{\alpha\beta},
\label{eq:10_norm}
\end{equation}
with analogous definitions for the conjugate multiplets.

\subsection{$D$-term scalar potential}

The global supersymmetric $D$-term potential associated with
$SU(5)\times U(1)_X$ is
\begin{equation}
V_D
=
\frac{g_5^2}{2}
\sum_{A=1}^{24}
\left(D_5^A\right)^2
+
\frac{g_X^2}{2}D_X^2,
\label{eq:VD_general}
\end{equation}
where $g_5$ and $g_X$ are the corresponding gauge couplings. The $SU(5)$
auxiliary fields are
\begin{align}
D_5^A ={}&
10_H^\dagger T_{10}^A10_H
+
\overline{10}_H^\dagger
T_{\overline{10}}^A\overline{10}_H
\nonumber\\
&+
\sum_a10_a^\dagger T_{10}^A10_a
+
\overline{10}_V^\dagger
T_{\overline{10}}^A\overline{10}_V,
\end{align}
where
\begin{equation}
T_{\overline{10}}^A
=
-\left(T_{10}^A\right)^\ast,
\end{equation}
are the generators of $SU(5)$. Using
\begin{equation}
X(10_H)=X(10_a)=+1,
\qquad
X(\overline{10}_H)=X(\overline{10}_V)=-1,
\end{equation}
the $U(1)_X$ auxiliary field is
\begin{equation}
D_X
=
\|10_H\|^2
+
\sum_a\|10_a\|^2
-
\|\overline{10}_H\|^2
-
\|\overline{10}_V\|^2.
\end{equation}
The complete $D$-term contribution may therefore be written as
\begin{align}
V_D ={}&
\frac{g_5^2}{2}
\sum_{A=1}^{24}
\Bigg[
10_H^\dagger T_{10}^A10_H
+
\sum_a10_a^\dagger T_{10}^A10_a
\nonumber\\
&\hspace{2.2cm}
-
\overline{10}_H^\dagger
\left(T_{10}^A\right)^\ast
\overline{10}_H
-
\overline{10}_V^\dagger
\left(T_{10}^A\right)^\ast
\overline{10}_V
\Bigg]^2
\nonumber\\
&+
\frac{g_X^2}{2}
\left(
\|10_H\|^2
+
\sum_a\|10_a\|^2
-
\|\overline{10}_H\|^2
-
\|\overline{10}_V\|^2
\right)^2.
\label{eq:VD_full}
\end{align}

\subsection{Right-handed neutrino field directions}

We now restrict the scalar potential to the electrically neutral field
subspace relevant for inflation. Using the gauge freedom, the GUT Higgs
multiplets are aligned along the Standard-Model-singlet directions
$N_H^c$ and $\overline N_H^c$. All gauge non-singlet components of the
matter and vector-like multiplets are set to zero, leaving only
\begin{equation}
N_a^c,\quad
\overline N_V^c,\quad
N_H^c,\quad
\overline N_H^c,\quad
S.
\end{equation}

Along this subspace, the two higher-dimensional interactions in
Eq.~(\ref{Winf}) reduce to the same field monomial. It is therefore convenient
to introduce the effective coupling
\begin{equation}
\alpha_a\equiv\beta_a+\gamma_a.
\label{eq:alpha_def}
\end{equation}
The $F$-term potential then becomes
\begin{align}
V_F^{\rm RHN} ={}&
\left|
\frac{\left(N_H^c\overline N_H^c\right)^2}{M_s^2}
-\mu_s^2
\right|^2
\nonumber\\
&+
\left|
\frac{2S}{M_s^2}
N_H^c\left(\overline N_H^c\right)^2
+
\frac{\overline N_H^c\overline N_V^c}{M_s}
\sum_a\alpha_aN_a^c
\right|^2
\nonumber\\
&+
\left|
\frac{2S}{M_s^2}
\left(N_H^c\right)^2\overline N_H^c
+
\frac{N_H^c\overline N_V^c}{M_s}
\sum_a\alpha_aN_a^c
\right|^2
\nonumber\\
&+
\frac{
|N_H^c\overline N_H^c|^2
|\overline N_V^c|^2
}{M_s^2}
\sum_a|\alpha_a|^2
\nonumber\\
&+
\frac{|N_H^c\overline N_H^c|^2}{M_s^2}
\left|
\sum_a\alpha_aN_a^c
\right|^2.
\label{eq:VF_RHN_compact}
\end{align}

For the neutral singlet directions, only the diagonal $SU(5)$ generator
contributes. Adopting the fundamental-representation convention
\begin{equation}
T_{24}^{(5)}
=
\frac{1}{2\sqrt{15}}
\operatorname{diag}(2,2,2,-3,-3),
\label{eq:T24}
\end{equation}
the $D$-term potential reduces to
\begin{align}
V_D^{\rm RHN}
={}&
\frac{1}{2}
\left(
\frac{3}{5}g_5^2+g_X^2
\right)
\Bigg(
|N_H^c|^2
+
\sum_a|N_a^c|^2
\nonumber\\
&\hspace{2.7cm}
-
|\overline N_H^c|^2
-
|\overline N_V^c|^2
\Bigg)^2.
\label{eq:VD_RHN}
\end{align}

\subsection{Inflationary trajectory}

As discussed in the next section, suitable non-minimal terms in the K\"ahler
potential stabilize the scalar component of $S$ at the origin. The
inflationary dynamics then involve the matter sneutrino fields
$N_a^c$ and $\overline N_V^c$, together with the Higgs waterfall fields
$N_H^c$ and $\overline N_H^c$.

During inflation, the waterfall fields are stabilized at the origin,
\begin{equation}
N_H^c=\overline N_H^c=0,
\label{eq:waterfall_origin}
\end{equation}
while the inflaton evolves along a $D$-flat direction satisfying
\begin{equation}
V_D^{\rm RHN}=0,
\qquad
\sum_a|N_a^c|^2
=
|\overline N_V^c|^2.
\label{eq:Dflat_general}
\end{equation}
Along this trajectory, the global supersymmetric $F$-term potential reduces
to the constant vacuum energy,
\begin{equation}
V_F=\mu_s^4,
\label{eq:global_flat_potential}
\end{equation}
which provides the energy density during inflation.

In general, inflation may proceed along any $D$-flat combination of the
right-handed sneutrino fields $N_a^c$ and the vector-like field
$\overline N_V^c$. For definiteness, we restrict the subsequent analysis to
the single-field trajectory
\begin{equation}
|N_1^c|
=
|\overline N_V^c|
\equiv\Phi,
\qquad
N_{a\neq1}^c=0.
\label{eq:single_Dflat}
\end{equation}
The vector-like multiplet $\overline{10}_V$ is therefore essential for
constructing a $D$-flat inflationary direction involving the matter
sneutrino. The fields orthogonal to this trajectory are assumed to acquire
masses larger than the Hubble scale from the combined superpotential and
K\"ahler interactions and are consequently stabilized during inflation.

At the level of global supersymmetry, however, the potential in
Eq.~(\ref{eq:global_flat_potential}) is exactly flat along the inflationary
direction. Moreover, the waterfall fields remain stable at the origin, so
neither the slope required for slow-roll evolution nor the tachyonic
instability needed to terminate inflation is generated. In the following
section, we include supergravity effects through a non-minimal K\"ahler
potential. These corrections stabilize the singlet field, lift the flat
inflaton direction, and generate the waterfall instability that provides a
graceful exit from inflation.
\section{Inflationary Potential in Supergravity}
\label{sec:SUGRA_potential}
In this section, we incorporate supergravity (SUGRA)
effects through a non-minimal K\"ahler potential. These corrections stabilize
the gauge-singlet field $S$, lift the flat inflaton direction, and induce the
waterfall instability that provides a graceful exit from inflation. We also
show that radiative and soft supersymmetry-breaking corrections remain
subdominant throughout the parameter region of interest.

\subsection{Supergravity scalar potential and K\"ahler expansion}
\label{subsec:SUGRA_general}

The $F$-term scalar potential in $\mathcal N=1$ supergravity is given by
\begin{equation}
\label{eq:SUGRA_F_term_potential}
V_F
=
e^{K/m_P^2}
\left[
K^{i\bar j}
D_iW
\left(D_jW\right)^{\!*}
-
\frac{3}{m_P^2}|W|^2
\right],
\end{equation}
where
\begin{equation}
D_i W = \partial_i W \;+\; \frac{W}{m_P^2}\, \partial_i K,
\,
K_{i\bar{j}} = \partial_i \partial_{\bar{j}} K,
\,
K^{i\bar{j}} = (K_{i\bar{j}})^{-1}.
\end{equation}

Restricting the superpotential in Eq.~(\ref{Winf}) to the
right-handed-neutrino and GUT-Higgs components relevant for inflation, we obtain
\begin{align}
W_{\rm inf} ={}&
S\left[
\frac{\left(N_H^c\overline N_H^c\right)^2}{M_s^2}
-\mu_s^2
\right]
\nonumber  \\
&+
\frac{\gamma_a}{M_s}
\left(N_a^c\overline N_V^c\right)
\left(N_H^c\overline N_H^c\right)
\nonumber\\
&+
\frac{\beta_a}{M_s}
\left(N_a^c\overline N_H^c\right)
\left(\overline N_V^c N_H^c\right),
\label{eq:Winf_RHN}
\end{align}
where summation over the index $a$ is understood.

We consider a non-minimal K\"ahler potential containing all operators relevant
to the inflationary dynamics up to the required order in $m_P^{-1}$:
\begin{align}
K \supset{}&
|S|^2
+|N_H^c|^2
+|\overline N_H^c|^2
+\sum_a|N_a^c|^2
+|\overline N_V^c|^2
\nonumber  \\
&+
\frac{\kappa_S}{4m_P^2}|S|^4
+
\frac{\kappa_H}{4m_P^2}|N_H^c|^4
+
\frac{\kappa_{\bar H}}{4m_P^2}
|\overline N_H^c|^4
\nonumber \\
&+
\frac{\kappa_{\bar V}}{4m_P^2}
|\overline N_V^c|^4
+
\frac{1}{4m_P^2}
\sum_{a,b}\kappa_{ab}
|N_a^c|^2|N_b^c|^2
\nonumber \displaybreak[3]  \\
&+
\frac{\kappa_{SH}}{m_P^2}
|S|^2|N_H^c|^2
+
\frac{\kappa_{S\bar H}}{m_P^2}
|S|^2|\overline N_H^c|^2
\nonumber\\
&+
\frac{1}{m_P^2}
\sum_a\kappa_{Sa}|S|^2|N_a^c|^2
+
\frac{\kappa_{S\bar V}}{m_P^2}
|S|^2|\overline N_V^c|^2
\nonumber\\
&+
\frac{\kappa_{H\bar H}}{m_P^2}
|N_H^c|^2|\overline N_H^c|^2
+
\frac{1}{m_P^2}
\sum_a\kappa_{Ha}|N_H^c|^2|N_a^c|^2
\nonumber  \\
&+
\frac{\kappa_{H\bar V}}{m_P^2}
|N_H^c|^2|\overline N_V^c|^2
+
\frac{1}{m_P^2}
\sum_a\kappa_{\bar Ha}
|\overline N_H^c|^2|N_a^c|^2
\nonumber\\
&+
\frac{\kappa_{\bar H\bar V}}{m_P^2}
|\overline N_H^c|^2|\overline N_V^c|^2
+
\frac{1}{m_P^2}
\sum_a\kappa_{a\bar V}
|N_a^c|^2|\overline N_V^c|^2
\nonumber\\
&+
\frac{\eta_H}{2m_P^2}
\left[
\left(N_H^c\overline N_H^c\right)^2
+\mathrm{h.c.}
\right]
\nonumber\\
&+
\frac{1}{2m_P^2}
\sum_a\eta_a
\left[
\left(N_a^c\overline N_V^c\right)^2
+\mathrm{h.c.}
\right]
+
\Delta K^{(6)}_{S\Phi}.
\label{eq:Kahler_potential}
\end{align}

Only a restricted set of sixth-order terms contributes to the inflaton
potential at order $\mathcal O(m_P^{-4})$. These terms are collected in
\begin{align}
\Delta K^{(6)}_{S\Phi}
={}&
\frac{|S|^2}{4m_P^4}
\sum_{a,b}\kappa_{Sab}
|N_a^c|^2|N_b^c|^2
+
\frac{\kappa_{S\bar V\bar V}}{4m_P^4}
|S|^2|\overline N_V^c|^4
\nonumber\\
&+
\frac{|S|^2}{m_P^4}
\sum_a\kappa_{Sa\bar V}
|N_a^c|^2|\overline N_V^c|^2
\nonumber\\
&+
\frac{|S|^2}{2m_P^4}
\sum_a\eta_{Sa}
\left[
\left(N_a^c\overline N_V^c\right)^2
+\mathrm{h.c.}
\right].
\label{eq:Kahler_sixth}
\end{align}

\subsection{Stabilization of the singlet and waterfall fields}
\label{subsec:stabilization}

Expanding the supergravity potential to leading order in $m_P^{-2}$, the
global supersymmetric potential is supplemented by Hubble-induced mass terms:
\begin{align}
V_F^{\rm SUGRA}\simeq{}&
\left|
\frac{\left(N_H^c\overline N_H^c\right)^2}{M_s^2}
-\mu_s^2
\right|^2
\Bigg[
1-\kappa_S\frac{|S|^2}{m_P^2}
\nonumber\\
&+
\left(1-\kappa_{SH}\right)
\frac{|N_H^c|^2}{m_P^2}
+
\left(1-\kappa_{S\bar H}\right)
\frac{|\overline N_H^c|^2}{m_P^2}
\nonumber\\
&+
\sum_a
\left(1-\kappa_{Sa}\right)
\frac{|N_a^c|^2}{m_P^2}
+
\left(1-\kappa_{S\bar V}\right)
\frac{|\overline N_V^c|^2}{m_P^2}
\Bigg]
\nonumber \displaybreak[3] \\
&+
\left|
\frac{2S}{M_s^2}
N_H^c(\overline N_H^c)^2
+
\frac{\overline {N}_H^c\overline N_V^c}{M_s}
\sum_a\alpha_aN_a^c
\right|^2
\nonumber\\
&+
\left|
\frac{2S}{M_s^2}
({N}_H^c)^2\overline {N}_H^c
+
\frac{N_H^c\overline N_V^c}{M_s}
\sum_a\alpha_aN_a^c
\right|^2
\nonumber\\
&+
\frac{|{N}_H^c\overline {N}_H^c|^2
|\overline N_V^c|^2}{M_s^2}
\sum_a|\alpha_a|^2
\nonumber\\
&+
\frac{|{N}_H^c\overline {N}_H^c|^2}{M_s^2}
\left|
\sum_a\alpha_aN_a^c
\right|^2
+
\mathcal O(m_P^{-4}).
\label{eq:VF_SUGRA_RHN}
\end{align}

The quartic K\"ahler correction proportional to $\kappa_S$ generates a
Hubble-induced mass for the singlet field,
\begin{equation}
m_S^2
\simeq
-3\kappa_S\mathcal H^2,
\qquad
\mathcal H^2
\simeq
\frac{\mu_s^4}{3m_P^2},
\end{equation}
where $\mathcal H$ denotes the Hubble parameter during inflation. For
$\kappa_S\lesssim-1/3$, the singlet is heavier than the Hubble scale and is
rapidly driven to its minimum at
$S=0$.
It therefore remains effectively frozen throughout inflation.

The mixed K\"ahler interactions involving $S$ and the GUT Higgs fields generate
Hubble-induced contributions to the waterfall masses. Expanding about the
inflationary trajectory
\begin{equation}
S=N_H^c=\overline N_H^c=0,
\end{equation}
one finds
\begin{equation}
V_{\rm SUGRA}
\supset
\frac{\mu_s^4}{m_P^2}
\left[
(1-\kappa_{SH})|{N}_H^c|^2
+
(1-\kappa_{S\bar H})|\overline {N}_H^c|^2
\right].
\label{eq:VHmassSUGRA}
\end{equation}
In contrast, the global supersymmetric interactions provide positive,
inflaton-dependent contributions,
\begin{equation}
V_{\rm global}
\supset
\frac{|\alpha_1|^2\Phi^4}{M_s^2}
\left(
|{N}_H^c|^2
+
|\overline {N}_H^c|^2
\right),
\label{eq:VHmassGlobal}
\end{equation}
where
$\Phi
\equiv
|N_1^c|
=
|\overline N_V^c|$
parameterizes the single-field $D$-flat inflaton trajectory.

Along the Higgs $D$-flat direction, we write
\begin{equation}
{N}_H^c=\overline {N}_H^c\equiv\frac{h}{2},
\end{equation}
where $h$ is the canonically normalized real waterfall mode. Assuming, for
simplicity,
\begin{equation}
\kappa_{SH}
=
\kappa_{S\bar H}
\equiv
\kappa_{\rm w},
\end{equation}
the effective waterfall mass is
\begin{equation}
m_h^2(\Phi)
\simeq
\frac{|\alpha_1|^2\Phi^4}{M_s^2}
+
(1-\kappa_{\rm w})
\frac{\mu_s^4}{m_P^2}.
\label{eq:mh}
\end{equation}

For $\kappa_{\rm w}>1$, the Hubble-induced contribution is negative. At large
inflaton values, however, the positive term proportional to $\Phi^4$
dominates and stabilizes the waterfall field at the origin. As $\Phi$
decreases during inflation, the waterfall mass vanishes at the critical value
defined by
\begin{equation}
m_h^2(\Phi_c)=0.
\end{equation}
Using $M^2=\mu_sM_s$, this gives
\begin{equation}
\Phi_c
=
\left[
\frac{\kappa_{\rm w}-1}{|\alpha_1|^2}
\frac{\mu_s^2}{m_P^2}
\right]^{1/4}
M.
\label{eq:Phi_c}
\end{equation}
For $\Phi>\Phi_c$, the waterfall direction remains stable. Once
$\Phi<\Phi_c$, its effective squared mass becomes negative and the Higgs fields
rapidly evolve toward the supersymmetric vacuum,
\begin{equation}
\langle {N}_H^c\overline {N}_H^c\rangle=M^2,
\end{equation}
thereby breaking $SU(5)\times U(1)_X$ to the Standard Model gauge group and
terminating inflation.

\subsection{Effective single-field inflationary potential}
\label{subsec:effective_inflaton}

Along the inflationary valley,
\begin{equation}
S=N_H^c=\overline N_H^c=0,
\qquad
|N_1^c|
=
|\overline N_V^c|
=
\frac{\phi}{2},
\end{equation}
where $\phi=2\Phi$ is the canonically normalized inflaton field. Expanding the
SUGRA potential to quartic order in $\phi/m_P$, the effective inflationary
potential takes the form
\begin{equation}
V(\phi)
=
\mu_s^4
\left[
1
+
\kappa_\phi\frac{\phi^2}{m_P^2}
+
\delta_\phi\frac{\phi^4}{m_P^4}
+
\mathcal O\!\left(\frac{\phi^6}{m_P^6}\right)
\right].
\label{eq:inflationpotential}
\end{equation}

The effective quadratic coefficient is
\begin{equation}
\kappa_\phi
=
\frac{1}{4}
\left(
2-\kappa_{S1}-\kappa_{S\bar V}
\right),
\label{eq:kphi}
\end{equation}
while the quartic coefficient is
\begin{align}
\delta_\phi
={}&
\frac{1}{16}
\Bigg[
2
+\frac{\kappa_{11}+\kappa_{\bar V}}{4}
+\kappa_{1\bar V}
+\eta_1
-2\left(\kappa_{S1}+\kappa_{S\bar V}\right)
\nonumber\\
&+
\left(\kappa_{S1}+\kappa_{S\bar V}\right)^2
-\frac{\kappa_{S11}+\kappa_{S\bar V\bar V}}{4}
-\kappa_{S1\bar V}
-\eta_{S1}
\Bigg].
\label{eq:dphi}
\end{align}

Throughout the analysis, we impose the nominal perturbative bound
\begin{equation}
|\kappa_i|,
\ |\eta_i|
\leq
\sqrt{4\pi}.
\label{eq:Kahler_bound}
\end{equation}
The corresponding ranges of the effective parameters are
\begin{equation}
\frac{1-2\sqrt{\pi}}{2}
\leq
\kappa_\phi
\leq
\frac{1+2\sqrt{\pi}}{2},
\end{equation}
and
\begin{equation}
\frac{1-10\sqrt{\pi}}{16}
\leq
\delta_\phi
\leq
\frac{2+18\sqrt{\pi}+16\pi}{16}.
\end{equation}
Numerically,
\begin{equation}
-1.27
\lesssim
\kappa_\phi
\lesssim
2.27,
\qquad
-1.05
\lesssim
\delta_\phi
\lesssim
5.26.
\end{equation}
Thus, values of $\delta_\phi$ close to $-1$, which are favored by the
inflationary analysis presented below, can be obtained without taking the
underlying K\"ahler coefficients beyond the nominal perturbative regime.

\subsection{Subleading corrections}
\label{subsec:subleading_corrections}

We now estimate the one-loop radiative and soft supersymmetry-breaking
corrections and show that both remain negligible compared with the tree-level
SUGRA potential.

\subsubsection{Coleman--Weinberg correction}

Along the single-field $D$-flat trajectory,
\begin{equation}
|N_1^c|
=
|\overline N_V^c|
=
\frac{\phi}{2},
\qquad
N_{a\neq 1}^c=0,
\end{equation}
the relevant waterfall interaction is
\begin{equation}
W_{\rm inf}
\supset
\frac{\alpha_1}{M_s}
N_1^c\overline N_V^c
N_H^c\overline N_H^c.
\end{equation}
The fermionic components of $N_H^c$ and $\overline N_H^c$ form a Dirac
fermion with inflaton-dependent mass
\begin{equation}
m_F(\phi)
=
\frac{|\alpha_1|}{M_s}\Phi^2
=
\frac{|\alpha_1|}{4M_s}\phi^2.
\label{eq:waterfall-fermion-mass}
\end{equation}

The corresponding scalar fields receive the same supersymmetric mass together
with Hubble-induced mass shifts:
\begin{align}
m_H^2(\phi)
&=
m_F^2(\phi)+\Delta_H,
\nonumber\\
m_{\bar H}^2(\phi)
&=
m_F^2(\phi)+\Delta_{\bar H},
\label{eq:waterfall-boson-masses}
\end{align}
where
\begin{align}
\Delta_H
&=
(1-\kappa_{SH})
\frac{\mu_s^4}{m_P^2}
=
3(1-\kappa_{SH})\mathcal H^2,
\nonumber\\
\Delta_{\bar H}
&=
(1-\kappa_{S\bar H})
\frac{\mu_s^4}{m_P^2}
=
3(1-\kappa_{S\bar H})\mathcal H^2.
\label{eq:waterfall-splittings}
\end{align}
These terms lift the supersymmetric mass degeneracy between the scalar and
fermionic waterfall states.

The corresponding Coleman--Weinberg correction is
\begin{align}
V_{\rm CW}(\phi)
={}&
\frac{1}{32\pi^2}
\Bigg[
m_H^4(\phi)
\left(
\ln\frac{m_H^2(\phi)}{Q^2}
-\frac{3}{2}
\right)
\nonumber\\
&+
m_{\bar H}^4(\phi)
\left(
\ln\frac{m_{\bar H}^2(\phi)}{Q^2}
-\frac{3}{2}
\right)
\nonumber\\
&-
2m_F^4(\phi)
\left(
\ln\frac{m_F^2(\phi)}{Q^2}
-\frac{3}{2}
\right)
\Bigg],
\label{eq:CW-waterfall-general}
\end{align}
where $Q$ is the renormalization scale.
For
$\kappa_{SH}
=
\kappa_{S\bar H}
=
\kappa_{\rm w}$,
and choosing
\begin{equation}
Q
=
\frac{|\alpha_1|\phi_c^2}{4M_s},
\end{equation}
the correction normalized to $V_0=\mu_s^4$ may be written as
\begin{align}
\frac{V_{\rm CW}}{V_0}
={}&
\frac{(\kappa_{\rm w}-1)^2}{16\pi^2}
\frac{\mu_s^4}{m_P^4}
\Bigg[
(x-1)^2
\left(
\ln|x-1|-\frac{3}{2}
\right)
\nonumber\\
&\hspace{3cm}
-
x^2
\left(
\ln x-\frac{3}{2}
\right)
\Bigg].
\label{eq:CW_normalized}
\end{align}
where $x\equiv
\left(\phi / \phi_c \right)^4$.

The Coleman--Weinberg contribution is suppressed both by the usual loop factor
$1/(16\pi^2)$ and by the additional factor $(\mu_s/m_P)^4$. We have verified
numerically throughout the parameter region of interest that it remains much
smaller than the tree-level SUGRA potential. It therefore has no appreciable
effect on the inflationary evolution or on the predicted observables. 
It is worth
noting, however, that the situation can be markedly different in
non-supersymmetric hybrid inflation, where one-loop Coleman--Weinberg
corrections induced by inflaton couplings to fields involved in reheating can
significantly modify the inflationary potential and bring the model predictions
into better agreement with the Planck and ACT constraints
~\cite{Ahmed:2025sfm,Ahmed:2026agd}.

\subsubsection{Soft supersymmetry-breaking correction}

Soft supersymmetry-breaking effects are likewise negligible during inflation.
Along the inflationary trajectory,
\begin{equation}
W_{\rm inf}=0,
\qquad
D_SW_{\rm inf}=-\mu_s^2,
\end{equation}
so the vacuum energy originates from the nonvanishing singlet $F$-term. Since
the superpotential itself vanishes, the conventional soft linear term does not
contribute along the inflationary valley.

The remaining soft scalar masses contribute terms of order
$m_{3/2}^2\phi^2$. For a representative gravitino mass
$m_{3/2}\sim\mathcal O(10~{\rm TeV})$, one has
\begin{equation}
m_{3/2}^2
\ll
\mathcal H^2
\simeq
\frac{\mu_s^4}{3m_P^2}.
\end{equation}
Their contribution is therefore strongly suppressed relative to the
Hubble-induced SUGRA terms. We have explicitly checked that including these
soft masses produces no considerable change in the inflationary trajectory or
the resulting observables. Accordingly, both radiative and soft
supersymmetry-breaking corrections may be safely neglected in the subsequent
analysis.

\section{Inflationary Dynamics and Observables}
\label{sec:inflation-dynamics}

\subsection{Slow-Roll Formalism}

The inflationary predictions are computed using the standard slow-roll
formalism. The slow-roll parameters are defined by
\begin{align}
\epsilon &=
\frac{m_P^2}{2}
\left(\frac{V'}{V}\right)^2,
&
\eta &=
m_P^2\frac{V''}{V},
&
\zeta^2 &=
m_P^4\frac{V'V'''}{V^2},
\end{align}
where primes denote derivatives with respect to the canonically normalized
inflaton field $\phi$.

To leading order in the slow-roll expansion, the scalar spectral index,
tensor-to-scalar ratio, and running of the scalar spectral index are given by
\begin{align}
n_s
&=
1-6\epsilon_0+2\eta_0,
\label{eq:ns-general}
\\
r
&=
16\epsilon_0,
\label{eq:r-general}
\\
\alpha_s
\equiv
\frac{dn_s}{d\ln k}
&=
16\epsilon_0\eta_0
-24\epsilon_0^2
-2\zeta_0^2,
\label{eq:alphas-general}
\end{align}
where the subscript ``0'' denotes evaluation at the field value
$\phi_0\equiv\phi(k_0)$ corresponding to the pivot scale
$k_0=0.05~{\rm Mpc}^{-1}$.

The amplitude of the primordial curvature perturbation at the pivot scale is
\begin{equation}
A_s(k_0)
=
\frac{V(\phi_0)}
{24\pi^2m_P^4\,\epsilon_0},
\label{eq:As}
\end{equation}
which is normalized to the observed value
$A_s^{\rm obs} = 2.137\times10^{-9}$.

The number of $e$-folds between horizon exit and the end of inflation is
\begin{equation}
N_0
=
\frac{1}{m_P^2}
\int_{\phi_e}^{\phi_0}
\frac{V}{V'}\,d\phi,
\label{eq:N0}
\end{equation}
where the end of inflation is identified with the waterfall transition,
$\phi_e=\phi_c$, corresponding to the SUGRA-induced tachyonic instability
discussed in Eq.~(\ref{eq:Phi_c}).

\subsection{Slow-Roll Analysis of the Inflationary Potential}

The inflationary dynamics are governed by the effective single-field
potential
\begin{equation}
V(\phi)
=
\mu_s^4
\left(
1+\kappa_\phi\frac{\phi^2}{m_P^2}
+\delta_\phi\frac{\phi^4}{m_P^4}
\right).
\end{equation}
It is convenient to introduce the dimensionless inflaton field
\begin{equation}
u \equiv\frac{\phi}{m_P},
\qquad
f(u)\equiv
1+\kappa_\phi u^2+\delta_\phi u^4,
\end{equation}
so that
\begin{equation}
V(\phi)=\mu_s^4 f(u).
\end{equation}

The number of $e$-folds between horizon exit and the end of inflation is given by
\begin{equation}
N_0
=
\int_{u_e}^{u_0}
\frac{f(u)}
{2u(\kappa_\phi+2\delta_\phi u^2)}
\,du,
\label{eq:efolds-integral}
\end{equation}
where $u_0=\phi_0/m_P$ denotes the field value at horizon exit.
Since inflation ends through the waterfall transition,
\begin{equation}
u_e=u_c\equiv\frac{\phi_c}{m_P},
\end{equation}
provided that the slow-roll conditions remain valid up to the critical point.

For nonzero $\kappa_\phi$ and $\delta_\phi$, the above integral admits the closed-form expression
\begin{align}
N_0
={}&
\frac{1}{4}
\Bigg[
\frac{1}{\kappa_\phi}\ln u^2
+\left(
\frac{\kappa_\phi}{4\delta_\phi}
-\frac{1}{\kappa_\phi}
\right)
\ln\left(
\kappa_\phi+2\delta_\phi u^2
\right)
+\frac{u^2}{2}
\Bigg]_{u_e}^{u_0}.
\label{eq:efolds-analytic}
\end{align}
For specified values of
$N_0$,
$\kappa_\phi$,
$\delta_\phi$,
and
$u_c$,
Eq.~(\ref{eq:efolds-analytic})
determines the horizon-exit field value
$u_0$.

Substituting the exact slow-roll parameters into
Eqs.~(\ref{eq:ns-general})--(\ref{eq:alphas-general}),
the inflationary observables are given by
\begin{align}
n_s
={}&
1
-
\frac{
12u_0^2
\left(\kappa_\phi+2\delta_\phi u_0^2\right)^2
}{
f^2(u_0)
}
+
\frac{
4\kappa_\phi+24\delta_\phi u_0^2
}{
f(u_0)
},
\label{eq:ns-exact}
\\
r
={}&
\frac{
32u_0^2
\left(\kappa_\phi+2\delta_\phi u_0^2\right)^2
}{
f^2(u_0)
},
\label{eq:r-exact}
\end{align}
and
\begin{align}
\alpha_s
={}&
\frac{
32u_0^2
\left(\kappa_\phi+2\delta_\phi u_0^2\right)^2
\left(2\kappa_\phi+12\delta_\phi u_0^2\right)
}{
f^3(u_0)
}
\nonumber\\
&-
\frac{
96u_0^4
\left(\kappa_\phi+2\delta_\phi u_0^2\right)^4
}{
f^4(u_0)
}
\nonumber\\
&-
\frac{
96\delta_\phi u_0^2
\left(\kappa_\phi+2\delta_\phi u_0^2\right)
}{
f^2(u_0)
}.
\label{eq:alphas-exact}
\end{align}

Although the exact expressions are used throughout the numerical analysis,
considerable insight can be gained in the vacuum-energy-dominated regime,
$f(u_0)\simeq1$,
which is realized over the phenomenologically relevant parameter space.
In this limit, the inflationary observables simplify to
\begin{align}
n_s
\simeq{}&
1
-12u_0^2
\left(\kappa_\phi+2\delta_\phi u_0^2\right)^2
+4\kappa_\phi
+24\delta_\phi u_0^2,
\label{eq:ns-approx}
\\
r
\simeq{}&
32u_0^2
\left(\kappa_\phi+2\delta_\phi u_0^2\right)^2,
\label{eq:r-approx}
\\
\alpha_s
\simeq{}&
32u_0^2
\left(\kappa_\phi+2\delta_\phi u_0^2\right)^2
\left(
2\kappa_\phi+12\delta_\phi u_0^2
\right)
\nonumber\\
&-
96u_0^4
\left(\kappa_\phi+2\delta_\phi u_0^2\right)^4
\nonumber\\
&-
96\delta_\phi u_0^2
\left(\kappa_\phi+2\delta_\phi u_0^2\right).
\label{eq:alphas-approx}
\end{align}
Furthermore, throughout the viable parameter space one typically finds
$\epsilon_0\ll|\eta_0|$.
The scalar spectral index is therefore well approximated by
\begin{equation}
n_s
\simeq
1+4\kappa_\phi+24\delta_\phi u_0^2.
\label{eq:ns-leading}
\end{equation}
This expression clearly illustrates the interplay between the quadratic and
quartic SUGRA corrections.
For
$\kappa_\phi>0$
and
$\delta_\phi<0$,
the negative quartic contribution can dominate at horizon exit,
naturally yielding the observed red-tilted scalar spectrum.

In the same approximation, the scalar amplitude reduces to
\begin{equation}
A_s(k_0)
\simeq
\frac{\mu_s^4}{
48\pi^2m_P^4
u_0^2
\left(\kappa_\phi+2\delta_\phi u_0^2\right)^2
}.
\label{eq:As-approx}
\end{equation}
The observed normalization of the scalar power spectrum then determines the
inflationary energy scale according to
\begin{equation}
\mu_s
\simeq
m_P
\left[
48\pi^2 A_s(k_0)
u_0^2
\left(
\kappa_\phi+2\delta_\phi u_0^2
\right)^2
\right]^{1/4}.
\label{eq:mus-normalization}
\end{equation}
Using the observed value
$A_s^{\rm obs}=2.137\times10^{-9}$,
the inflationary energy scale may be expressed directly in terms of the
tensor-to-scalar ratio as
\begin{equation}
\mu_s
\simeq
6 \times10^{15}\ {\rm GeV}
\left(
\frac{r}{10^{-3}}
\right)^{1/4}.
\end{equation}
This relation follows from the fixed normalization of the scalar power
spectrum and highlights the close connection between the inflationary energy
scale and primordial gravitational waves.
In particular, values
$r\gtrsim10^{-3}$,
which lie within the projected sensitivity of forthcoming CMB $B$-mode
polarization experiment such as LiteBIRD \cite{LiteBIRD:2023aov}, correspond to an
inflationary energy scale of order
$10^{16}\ {\rm GeV}$,
close to the characteristic GUT scale.
\section{Numerical Results} \label{sec:numerical_results}
\subsection{Viable Sign Choices for $\kappa_{\phi}$ and 
$\delta_{\phi}$}
The inflationary dynamics is largely governed by the signs of the
quadratic and quartic coefficients, $\kappa_\phi$ and $\delta_\phi$, in
Eq.~(\ref{eq:inflationpotential}). Four qualitatively distinct
possibilities arise:

\begin{enumerate}
\item[(i)] $\boldsymbol{\kappa_\phi>0}$ and
$\boldsymbol{\delta_\phi<0}$:

This is the phenomenologically viable case considered in this work. The
positive quadratic term provides a positive curvature near the origin; in contrast, the negative quartic term gradually flattens the potential at
larger field values, leading to a monotonic slow-roll trajectory toward
the critical value $\phi_c$. This structure naturally yields a red-tilted
scalar spectrum, $n_s<1$, with sub-Planckian inflaton values and is
consistent with the latest CMB observations.

\item[(ii)] $\boldsymbol{\kappa_\phi<0}$ and
$\boldsymbol{\delta_\phi<0}$:

In this case both terms drive the inflaton away from the origin.
Consequently, the inflaton rolls in the direction opposite to the
waterfall point rather than toward the critical value $\phi_c$, and the
waterfall transition cannot be realized. This branch therefore does not
lead to a viable hybrid inflationary scenario.

\item[(iii)] $\boldsymbol{\kappa_\phi>0}$ and
$\boldsymbol{\delta_\phi>0}$:

Here the potential becomes increasingly steep as the inflaton moves away
from the origin. As a result, the slow-roll parameters are typically too
large to generate a sufficiently red-tilted scalar spectrum for
sub-Planckian field values. Numerical scans confirm that this branch is
incompatible with the observed CMB constraints.

\item[(iv)] $\boldsymbol{\kappa_\phi<0}$ and
$\boldsymbol{\delta_\phi>0}$:

The potential develops a local minimum at
$\phi_{\min}/m_P=\sqrt{|\kappa_\phi|/(2\delta_\phi)}$.
For $\phi<\phi_{\min}$, a red-tilted spectrum may be obtained, but the
inflaton rolls toward larger field values, away from the waterfall
point, and eventually becomes trapped in the local minimum at
$\phi_{\min}$. On the outer branch, $\phi>\phi_{\min}$, the inflaton
can roll toward the critical point, but the positive curvature generally
produces a blue-tilted spectrum. Therefore, this sign choice does not yield a viable hybrid-inflation trajectory.
\end{enumerate}

Therefore, within the effective potential truncated at
$\mathcal{O}(\phi^4/m_P^4)$, only the parameter choice
\[
\kappa_\phi>0,
\qquad
\delta_\phi<0,
\]
leads to a phenomenologically viable realization of sub-Planckian
hybrid inflation.

\subsection{Parameter Reduction and Numerical Procedure}
The effective inflaton potential truncated at quartic order is specified by the four parameters
$\{\mu_s,M_s,\kappa_\phi,\delta_\phi\}$.
Although the critical inflaton value $\phi_c$ is, in general, an independent parameter, for simplicity we identify it with the flipped $SU(5)$ symmetry-breaking scale,
\begin{equation}
\phi_c=M=\sqrt{\mu_sM_s},
\end{equation}
so that the end of inflation is completely determined once $\mu_s$ and $M_s$ are specified.

To reduce the dimensionality of the parameter space, we impose the
following conditions:
\begin{enumerate}
\item The scalar amplitude is fixed to the observed value,
      $A_s=A_s^{\rm obs}$.
\item  Following the standard post-inflationary thermal history, the number
of $e$-folds between the horizon exit of the pivot scale and the end of
inflation is determined by the reheating temperature $T_r$ and the
inflationary energy scale $\mu_s$ according to~\cite{Liddle:2003as}
\begin{equation}
N_0 \simeq
53+\frac{1}{3}\ln\!\left(\frac{T_r}{10^9~\mathrm{GeV}}\right)
+\frac{2}{3}\ln\!\left(\frac{\mu_s}{10^{15}~\mathrm{GeV}}\right),
\label{eq:TRN0}
\end{equation}
where $\mu_s^4$ is the inflationary vacuum energy density. Throughout
this work we adopt the representative reheating temperature
$T_r=10^6~{\rm GeV}$, motivated by successful non-thermal leptogenesis
and the gravitino constraint discussed in
Sec.~\ref{sec:reheating}. 
\item Inflation ends at the waterfall transition,
$\phi_e=\phi_c=\sqrt{\mu_s M_s}$.
\item The scalar spectral index is fixed to its central observed value,
      $n_s=0.9734$.
\end{enumerate}

The normalization condition $A_s=A_s^{\rm obs}$ determines the overall
inflationary scale $\mu_s$; however, the conditions on $N_0$ and $n_s$
determine the horizon-exit field value $\phi_0$ together with one of the
two shape parameters $(\kappa_\phi,\delta_\phi)$. 
Consequently, the original four-dimensional parameter space is reduced to two independent parameters.
In our numerical analysis, we choose any two parameters as the independent input parameters, while the
others are determined numerically by
solving the above system of equations. 

For example, we perform a two-dimensional numerical scan over the
$(M_s,\kappa_\phi)$ parameter space. The results are displayed as a sequence of one-dimensional slices obtained by fixing $M_s$ at representative values while varying $\kappa_\phi$.
This presentation clearly isolates the impact of each parameter without sacrificing the generality of the scan.

\subsection{Theoretical and Observational Constraints}

In our numerical analysis, we impose a number of theoretical and phenomenological constraints in order to identify the physically viable region of parameter space. 
We first require the inflaton field value at horizon exit to remain sub-Planckian,
\begin{equation}
\phi_0 < m_P,
\end{equation}
so that the supergravity expansion remains under theoretical control.

The flipped $SU(5)$ symmetry-breaking scale is taken to satisfy
\begin{equation}
M < 5\times10^{17}~{\rm GeV},
\end{equation}
where the upper bound is imposed to keep the symmetry-breaking scale below the expected string scale, $M_{\text{string}} \simeq 5\times10^{17}~{\rm GeV}$. 

To ensure that inflation does not occur arbitrarily close to the waterfall transition, we require
\begin{equation}
\Delta x \equiv \frac{\phi_0 - \phi_c}{\phi_c} > 0.01,
\end{equation}
which guarantees that the observable inflationary epoch takes place sufficiently far from the critical point. 
Throughout the numerical analysis we further restrict the quartic coefficient to the phenomenologically relevant range
\begin{equation}
|\delta_\phi| \le 1,
\end{equation}
in accordance with the discussion in the previous section.

The above constraints define the boundary of the parameter space used throughout our numerical analysis. Within this allowed region, we further distinguish the parameter space according to whether the inflationary field value exceeds the cutoff scale $M_s$. In particular, we include the curve defined by $\phi_0=M_s$, which partitions the allowed parameter space into two distinct regions.

Our primary focus is the region satisfying
\begin{equation}
\phi_0 < M_s,
\end{equation}
where the effective superpotential remains under perturbative control, since higher-dimensional operators suppressed by powers of $M_s$ are expected to be negligible. In contrast, for $\phi_0>M_s$, additional non-renormalizable operators may become important unless their coefficients are significantly suppressed or forbidden by an underlying ultraviolet completion. 
For this reason, the region with $\phi_0>M_s$ is shown as the shaded gray
area in the figures. In this region, higher-dimensional operators
suppressed by powers of $M_s$ are no longer guaranteed to be negligible. The predictions shown there should therefore be
interpreted with caution.

The recent ACT DR6 analysis~\cite{AtacamaCosmologyTelescope:2025nti}, allowing for a scale-dependent primordial spectrum, finds
\begin{equation}
\alpha_s = 0.0062 \pm 0.0052,
\end{equation}
from the P--ACT--LB data combination. 
Although the current data do not provide statistically significant evidence for a nonzero running, the central value is positive.
Since the present model also predicts a positive running, we conservatively impose the $2\sigma$ upper limit,
\begin{equation}
\alpha_s \leq 0.0166,
\end{equation}
as one of the observational constraints defining the allowed parameter space. This condition is represented by the green boundary curve in Figs.~\ref{fig:kphi}--\ref{fig:alpha}, while Fig.~\ref{fig:alpha} shows contours of constant $\alpha_s$ within the viable region.

\subsection{Numerical Scan: Predictions Across the Allowed Region}
\begin{figure}[t]\centering
\includegraphics[width=\columnwidth,height=\textheight,keepaspectratio]{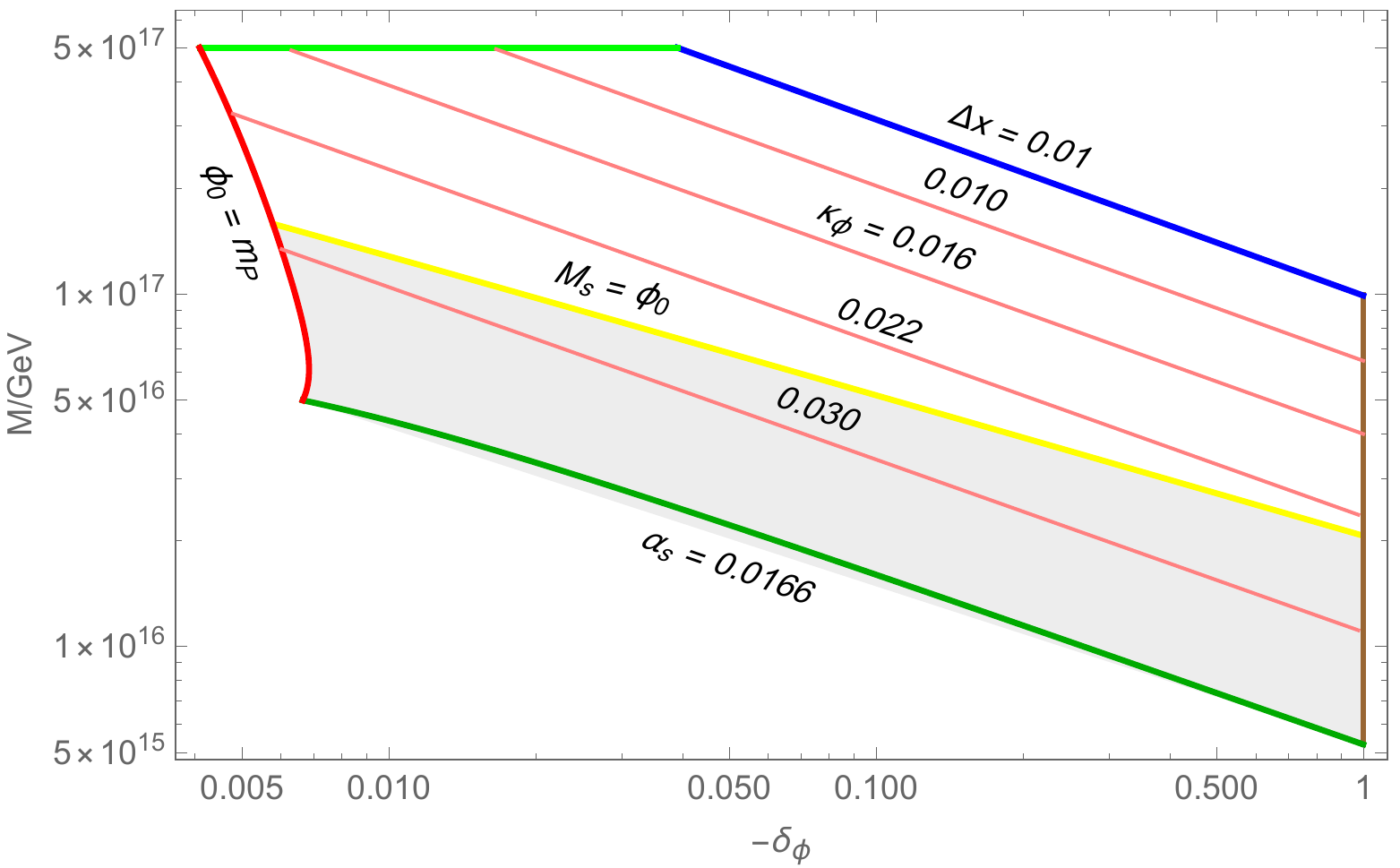}
\caption{
Allowed parameter space in the $M$--$|\delta_\phi|$ plane for
$\phi_c=M$, $n_s=0.9734$, and $T_r=10^6~\mathrm{GeV}$. 
The pink curves correspond to fixed values of $\kappa_{\phi}$ and illustrate the variation of
$\kappa_\phi$ ($0.0034 \lesssim \kappa_{\phi} \lesssim 0.04$) within the boundary region. The yellow curve, $M_s=\phi_0$, separates the viable
region $\phi_0<M_s$ from the grey-shaded region where the inflaton
field exceeds the cutoff scale. The red curve denotes
$\phi_0=m_P$, the blue curve corresponds to
$\Delta x = 0.01$, and the green lines indicate the
adopted bounds on $M$ and $|\delta_\phi|$.
}
\label{fig:kphi}
\end{figure}

\begin{figure}[t]
\centering
\includegraphics[width=\columnwidth,height=\textheight,keepaspectratio]{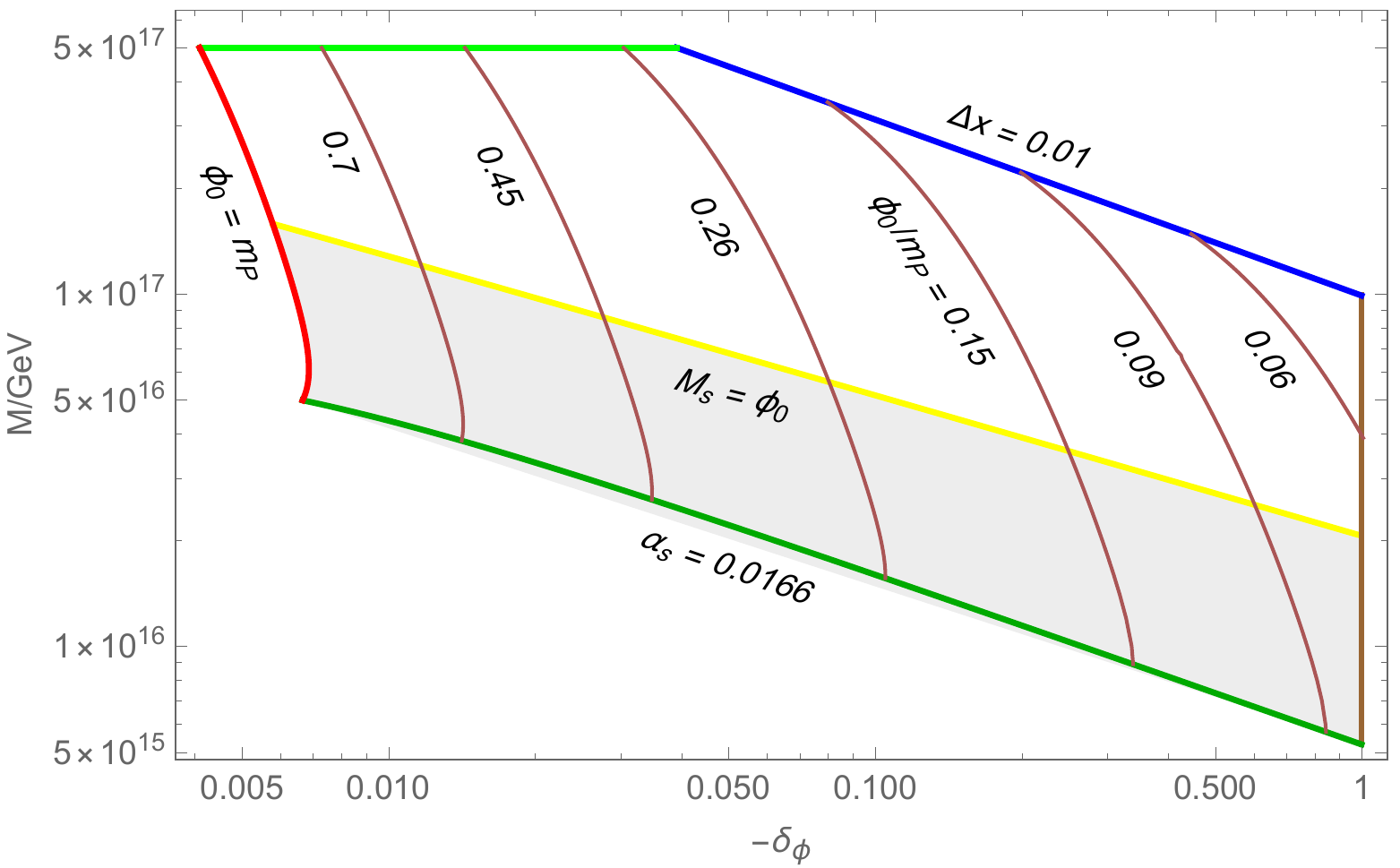}
\caption{
Allowed parameter space of Fig.~\ref{fig:kphi}
showing contours of constant inflaton field value, $\phi_0/m_P$, and displaying its variation, $0.04 \lesssim \phi_0/m_P \leq 1$, within the boundary region.
}
\label{fig:phi0}
\end{figure}

\begin{figure}[t]
\centering
\includegraphics[width=\columnwidth,height=\textheight,keepaspectratio]{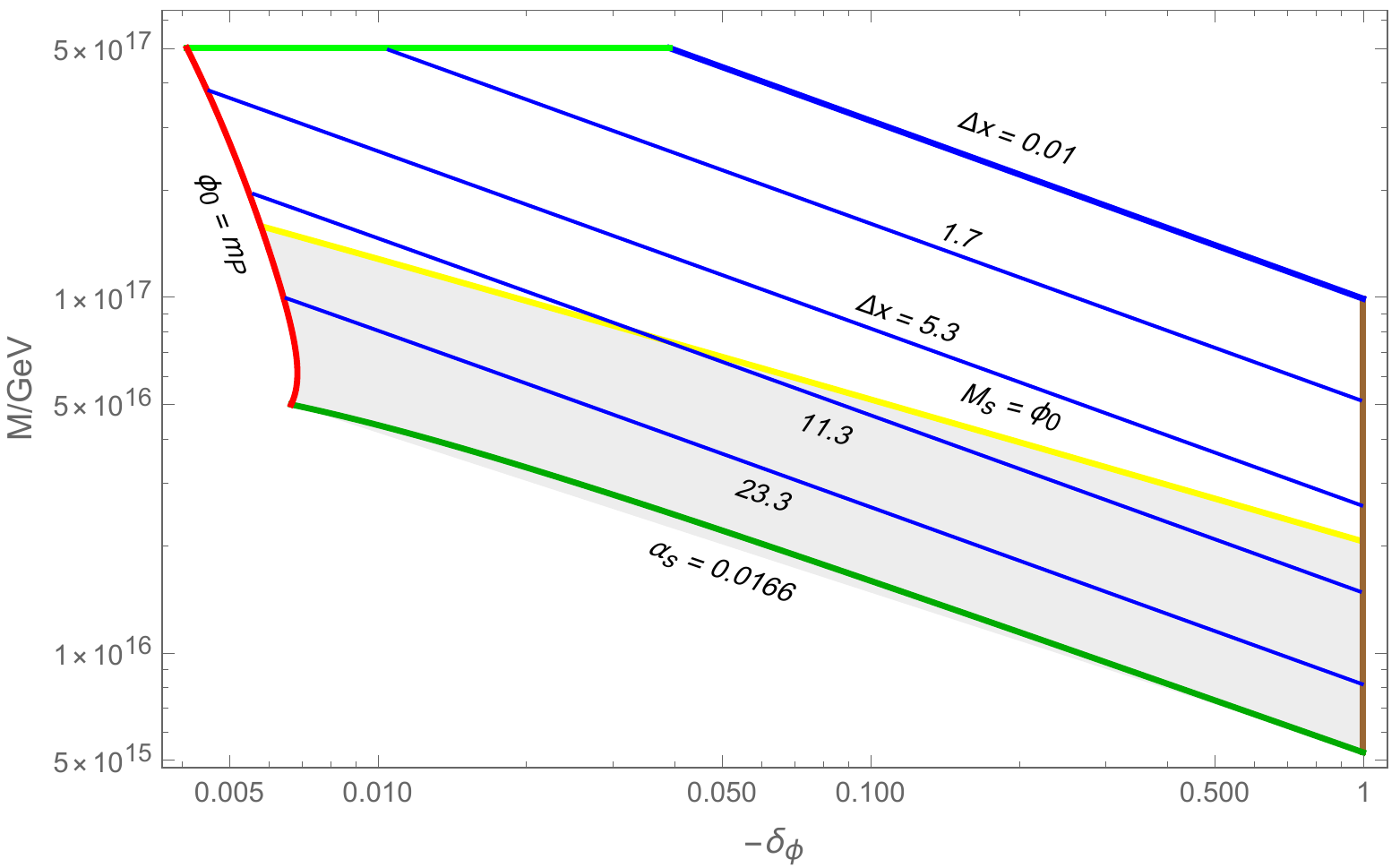}
\caption{
Allowed parameter space of Fig.~\ref{fig:kphi}
showing contours of constant  displacement parameter, $\Delta x$,  and displaying its variation, $0.01 \leq \Delta x \lesssim 48$, within the boundary region.
}
\label{fig:Deltax}
\end{figure}

\begin{figure}[t]
\centering
\includegraphics[width=\columnwidth,height=\textheight,keepaspectratio]{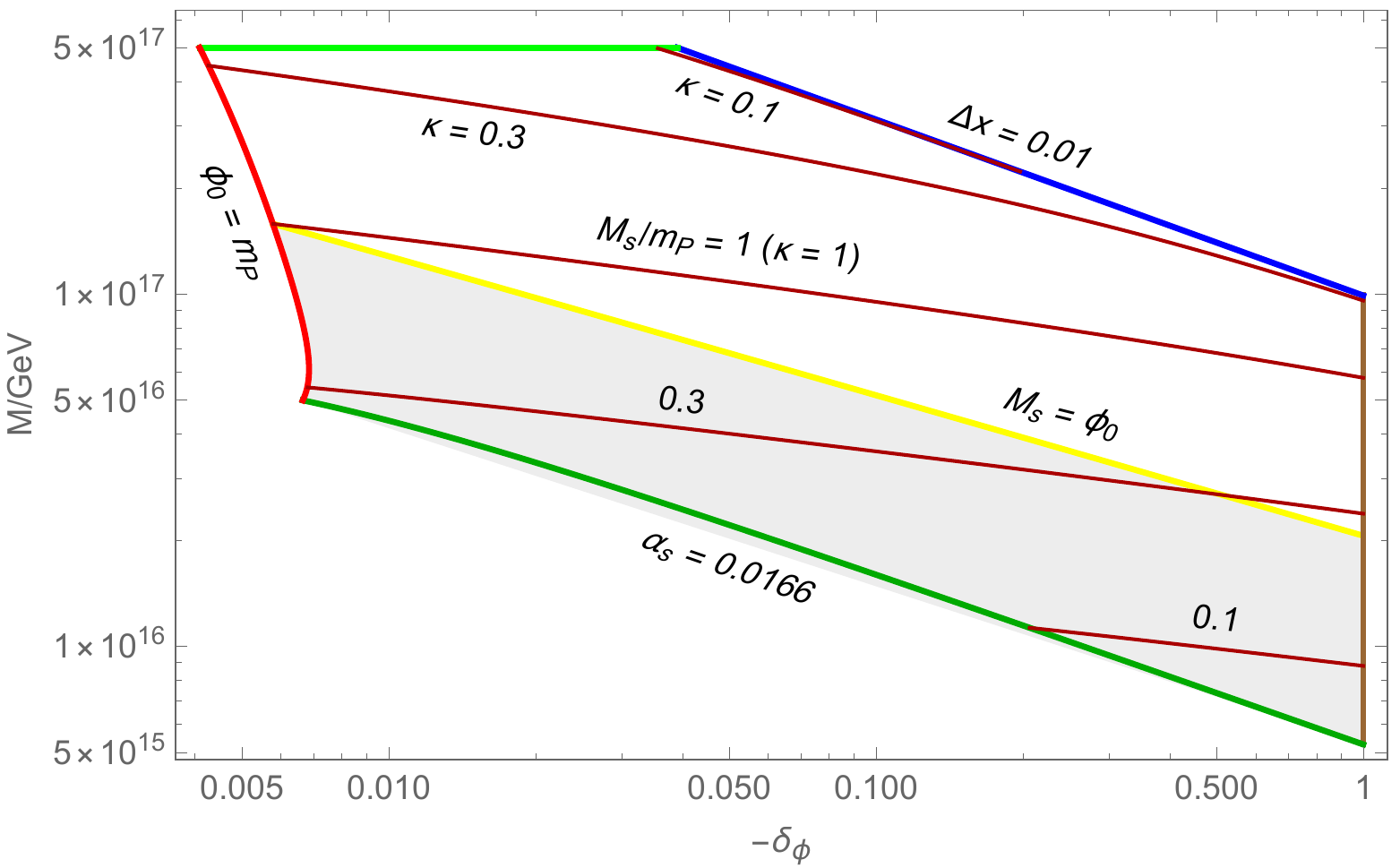}
\caption{
Allowed parameter space of Fig.~\ref{fig:kphi}
showing contours of constant  cutoff scale $M_s$, and displaying its variation, $0.003 \lesssim M_s/m_P \lesssim 1$, within the boundary region. For $M_s > m_P$, we redefine the effective cutoff scale as $M_s/\sqrt{\kappa}$ and fix $M_s = m_P$, allowing $\kappa$ to vary in the range $0.06 \lesssim \kappa \leq 1$.}
\label{fig:Ms}
\end{figure}

\begin{figure}[t]
\centering
\includegraphics[width=\columnwidth,height=\textheight,keepaspectratio]{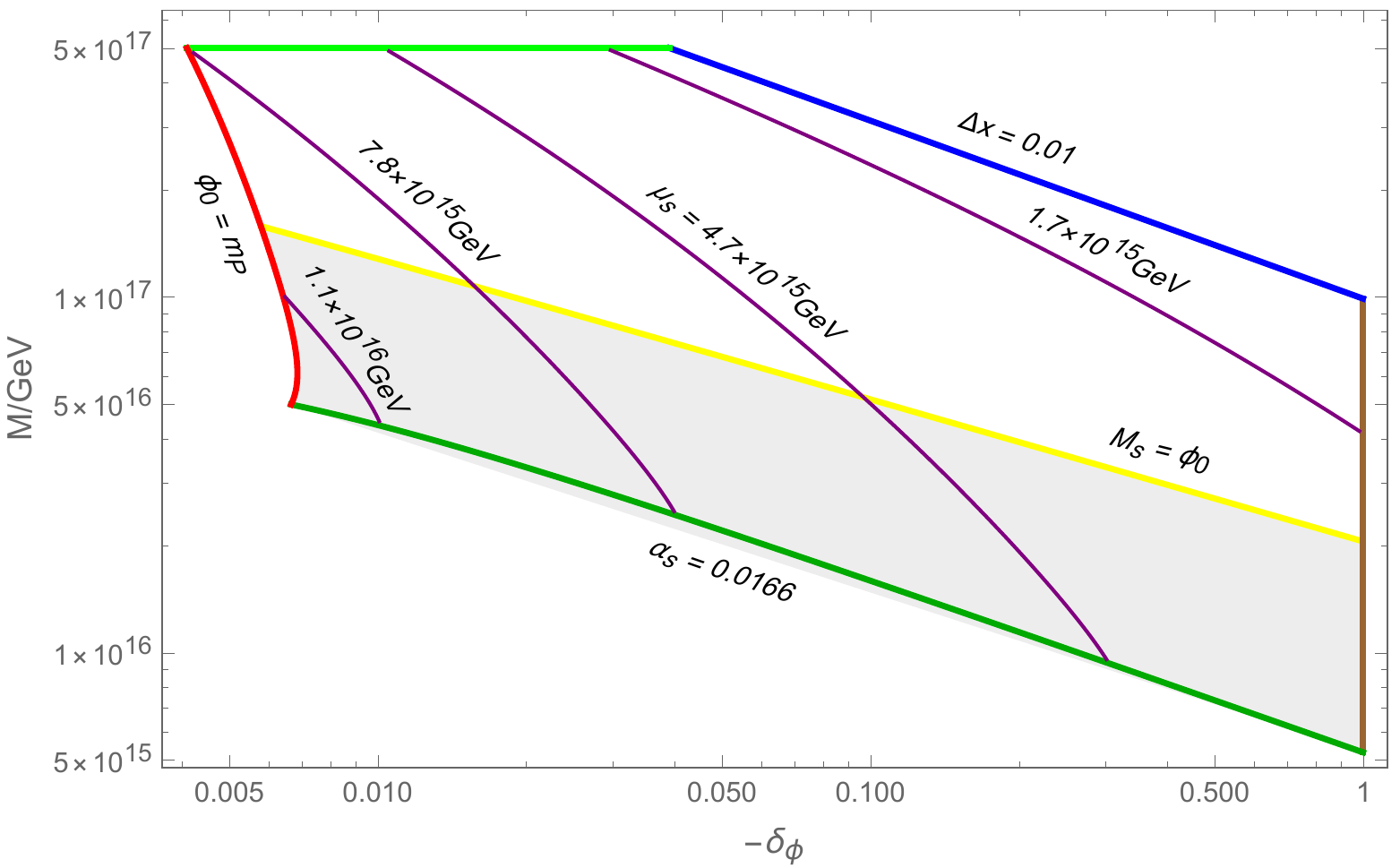}
\caption{
Same as Fig.~\ref{fig:kphi}, but showing contours of constant $\mu_s$, and displaying its variation, $ 10^{14}  \lesssim \mu_s/\text{GeV} \leq 10^{16}$, within the boundary region.
}
\label{fig:mus}
\end{figure}

\begin{figure}[t]
\centering
\includegraphics[width=\columnwidth,height=\textheight,keepaspectratio]{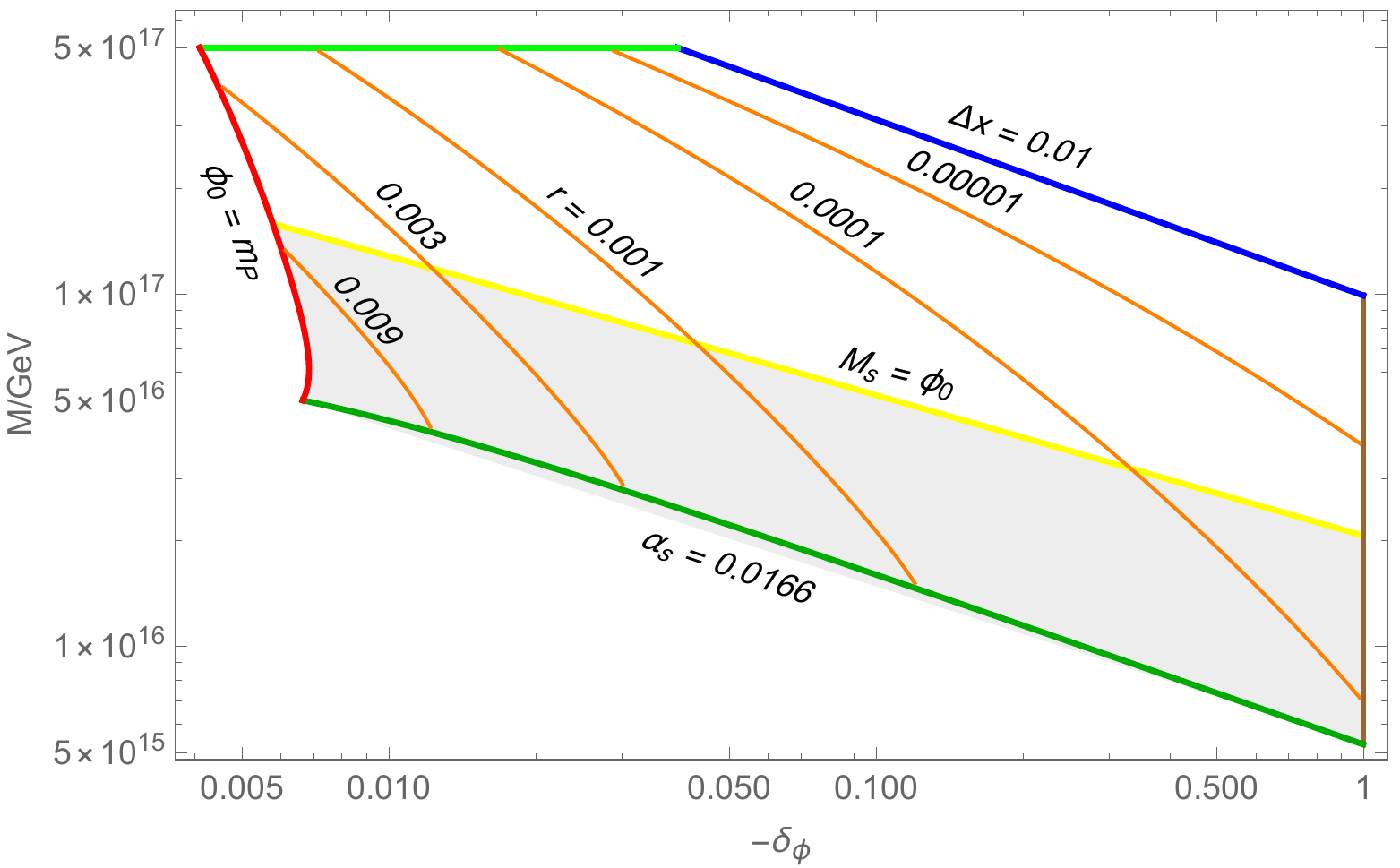}
\caption{
Allowed parameter space of Fig.~\ref{fig:kphi}
showing contours of constant  tensor-to-scalar ratio $r$, and displaying its variation, $ 3 \times 10^{-10} \lesssim r \lesssim 0.02$, within the boundary region.
}
\label{fig:r}
\end{figure}

\begin{figure}[t]
\centering
\includegraphics[width=\columnwidth,height=\textheight,keepaspectratio]{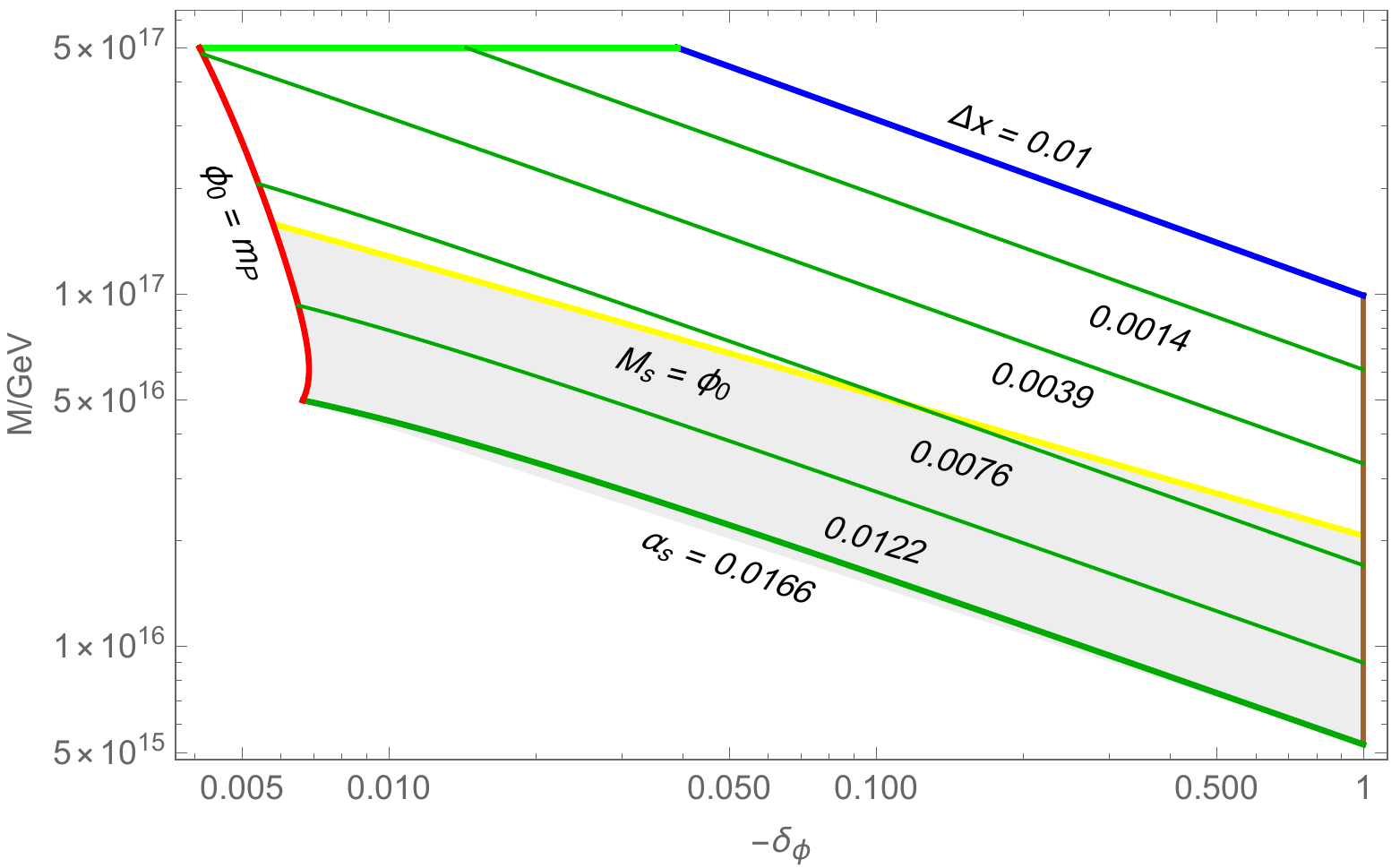}
\caption{
Allowed parameter space of Fig.~\ref{fig:kphi}
showing contours of constant running of the scalar spectral index, $\alpha_s$, 
and displaying its variation, $ 10^{-5}  \lesssim \alpha_s \leq 0.0166$, within the boundary region.
}
\label{fig:alpha}
\end{figure}

The numerical results presented below illustrate how the inflationary predictions vary across the phenomenologically allowed parameter space.
Figures~\ref{fig:kphi}--\ref{fig:alpha} display representative two-dimensional slices in the
$M$--$|\delta_\phi|$ plane together with contours of constant
$\kappa_\phi$,
$\phi_0$,
$\Delta x$,
$M_s$,
$\mu_s$,
$r$,
and
$\alpha_s$.

Figures~\ref{fig:kphi}--\ref{fig:alpha} display the allowed parameter space in the
$M$--$|\delta_\phi|$ plane for
$\phi_c=M$, $n_s=0.9734$, and $T_r=10^6~{\rm GeV}$.
The allowed region is bounded by the imposed limits on the symmetry-breaking scale $M$, the quartic coefficient $|\delta_\phi|$, the requirement $\phi_0<m_P$, the upper bound $\alpha_s < 0.0166$, and the lower bound $\Delta x>0.01$.
The curve $M_s=\phi_0$ separates the theoretically reliable region,
$\phi_0<M_s$, from the grey-shaded region in which the inflaton value at
horizon exit exceeds the cutoff scale. Over the complete parameter space, the ratio of the inflaton field value at horizon exit to the cutoff scale spans the range, 
$6\times10^{-4}
\lesssim
\phi_0 / M_s
\lesssim
26$.
Within the boundary region, the
effective quadratic coefficient varies approximately over
$0.0034\lesssim\kappa_\phi\lesssim0.04$, with larger values of
$\kappa_\phi$ generally obtained toward larger $|\delta_\phi|$ and smaller
$M$. The inflaton value at horizon exit spans
$0.04\lesssim\phi_0/m_P\leq1$, increasing toward the upper-right part of the
parameter space. Correspondingly, the displacement from the critical point,
$\Delta x$, ranges from values close to $0.01$ near the upper boundary to
values as large as $\mathcal O(10)$--$\mathcal O(50)$ toward the lower part of
the allowed region.

Having established the structure of the viable parameter space, we now examine how the underlying high-energy scales vary across this region.
Figure~\ref{fig:Ms} highlights one of the main differences between the
present analysis and earlier studies of K\"ahler-driven tribrid inflation.
Previous works typically fixed the cutoff scale to the reduced Planck mass,
$M_s=m_P$.
Here, we relax this assumption and treat $M_s$ as an independent parameter.
This considerably enlarges the viable parameter space while remaining
consistent with both theoretical consistency requirements and current
CMB observations.
The effective cutoff scale varies over a broad range,
approximately
$0.003\lesssim M_s/m_P \leq 1$.
For parameter choices that would formally give $M_s>m_P$, the cutoff is redefined as 
\begin{equation}
M_s \rightarrow M_s/\sqrt{\kappa},    
\end{equation}
with $M_s$ fixed at $m_P$ and 
$0.06\lesssim\kappa\leq1$.

Fig.~\ref{fig:mus} shows that the inflationary energy scale varies approximately
over the range
$10^{14}\lesssim \mu_s/{\rm GeV}\lesssim10^{16}$
and generally increases toward larger values of $M$.
For this range of inflationary scales, and adopting the representative
reheating temperature $T_r=10^6~{\rm GeV}$ discussed in
Sec.~\ref{sec:reheating}, Eq.~(\ref{eq:TRN0}) yields
\begin{equation}
49\lesssim N_0\lesssim52.
\end{equation}

Over most of the phenomenologically viable range of
$\delta_\phi$, the requirement
$\alpha_s\le0.0166$
imposes a lower bound
$M\gtrsim {\rm few}\times10^{16}$ GeV.
Consequently, the model naturally favors a flipped $SU(5)$
symmetry-breaking scale somewhat above the conventional MSSM
unification scale,
$M_{\rm GUT}\simeq2\times10^{16}$ GeV.
Only in the narrow corner of parameter space with
$|\delta_\phi|\rightarrow 1$
does $M$ approach, or fall slightly below,
$M_{\rm GUT}$.

We next turn to the inflationary observables ($r$ and $\alpha_s$) predicted within the
allowed parameter space.
The tensor-to-scalar ratio displayed in Fig.~\ref{fig:r} varies from
extremely small values,
$r\sim3\times10^{-10}$, up to values of order $10^{-2}$, with the largest
values occurring near the upper boundary of the allowed region. 
Although the predicted tensor-to-scalar ratio spans the wide range
$3\times10^{-10}\lesssim r\lesssim0.02$,
a substantial portion of the viable parameter space yields
$r\gtrsim10^{-3}$,
while remaining fully consistent with current observational bounds.
Such values are expected to be probed by forthcoming CMB $B$-mode polarization missions, including LiteBIRD \cite{LiteBIRD:2023aov}. Therefore, the model offers the realistic possibility of an observable primordial gravitational-wave signal in the near future.

Figures~\ref{fig:mus} and \ref{fig:r}
illustrate the close connection between the inflationary energy scale
and the tensor-to-scalar ratio.
The close resemblance is a direct consequence of the observed normalization of the scalar power spectrum. Since $A_s$ is fixed, the slow-roll relations imply
$V^{1/4}\propto r^{1/4}$.
In our model, $V\simeq\mu_s^4$ during inflation, so larger inflationary scales $\mu_s$ necessarily correspond to larger tensor-to-scalar ratios. This explains why the contours of constant $\mu_s$ and constant $r$ follow nearly identical trajectories in the allowed parameter space.

Figure~\ref{fig:alpha}
shows that the running remains positive throughout the allowed
parameter space and satisfies the adopted
$2\sigma$ ACT bound.
The largest values occur near the lower-right corner of the viable
region, where $|\delta_\phi|$ is relatively large and the
inflationary scale is lower.
Overall, the numerical analysis demonstrates that the proposed flipped
$SU(5)$ tribrid inflation model accommodates a broad range of symmetry-breaking, cutoff, and inflationary scales while remaining fully consistent with current theoretical requirements and CMB observations.
Furthermore, it predicts a potentially observable tensor-to-scalar ratio over a significant fraction of the viable parameter space, providing an important target for future CMB polarization experiments.

\section{Inflaton Decay, Reheating and Non-thermal Leptogenesis}
\label{sec:reheating}

\subsection{Heavy neutrino mass spectrum}

After the waterfall transition, the Standard-Model-singlet components
$N_H^c$ and $\overline N_H^c$ acquire vacuum expectation values of
magnitude $M$. The higher-dimensional operators involving
$N_a^c$, $\overline N_V^c$, $N_H^c$, and $\overline N_H^c$
in Eq.~(\ref{eq:Winf_RHN}) then generate the vector-like mixing terms,
\begin{equation}
W_{\rm mix}
=
\sum_a m_a N_a^c\overline N_V^c,
\qquad
m_a
=
\alpha_a\frac{M^2}{M_s}
=
\alpha_a\mu_s,
\end{equation}
where $\alpha_a=\beta_a+\gamma_a$. The quantities $m_a$ are entries of the
vector-like mixing vector rather than independent physical mass eigenvalues.

To generate Majorana masses for the right-handed neutrinos, we assume that
the $U(1)_R$ symmetry is broken by ultraviolet physics in such a way that the leading $R$-symmetry-breaking higher-dimensional operator \cite{Civiletti:2013cra,Rehman:2018gnr,Mehmood:2020irm,Abid:2021jvn,Ahmad:2025dds,Ahmad:2025mul}
\begin{equation}
W_{\rm Maj}
=
\frac{\lambda_{ab}}{2m_P}
\left(10_a\overline{10}_H\right)
\left(10_b\overline{10}_H\right),
\end{equation}
is allowed. This operator is invariant under the gauge and discrete
symmetries but carries an $R$ charge one unit below that required for the
superpotential. It may, for example, arise through an $R$-breaking spurion
$X_R$ satisfying $R(X_R)=1$,
\begin{equation}
W_{\rm Maj}
=
\frac{\lambda_{ab}}{2m_P}
\left(\frac{X_R}{M_{\rm UV}}\right)
\left(10_a\overline{10}_H\right)
\left(10_b\overline{10}_H\right).
\end{equation}
Once $X_R$ develops a vacuum expectation value, the suppression factor
$\langle X_R\rangle/M_{\rm UV}$ may be absorbed into the effective coupling
$\lambda_{ab}$.

The neutral heavy-fermion sector therefore contains both Majorana masses
for the right-handed neutrinos and Dirac-type mixings with the vector-like
state $\overline N_V^c$. We assume a hierarchical structure in which the
vector-like mixing is predominantly aligned with $N_V^c$,
\begin{equation}
|m_V| \gg |m_i|,
\qquad i=1,2,3.
\end{equation}
In addition, we assume that the off-diagonal Majorana mixings between the
ordinary right-handed neutrinos and the vector-like state are suppressed,
\begin{equation}
|\lambda_{iV}| \ll |\lambda_{VV}|,
\qquad i=1,2,3.
\end{equation}
These assumptions ensure that the vector-like sector is approximately
decoupled from the three ordinary right-handed neutrinos. Consequently,
$N_V^c$ and $\overline N_V^c$ combine predominantly into a superheavy
vector-like state, while the remaining $N_i^c$ are approximately Majorana
mass eigenstates governed by
\begin{equation}
(M_R)_{ij}
=
\lambda_{ij}\frac{M^2}{m_P},
\qquad i,j=1,2,3,
\end{equation}
and realize the conventional type-I seesaw mechanism.
We choose the
lightest of these states, $N_1^c$, and its scalar superpartner as the
right-handed (s)neutrino associated with the inflationary direction.
\subsection{Sneutrino inflaton direction and mass}

To realize sneutrino inflation while retaining a conventional type-I
seesaw sector, we choose the $D$-flat inflationary trajectory to be aligned
with the lightest ordinary right-handed sneutrino,
\begin{equation}
|N_1^c|
=
|\overline N_V^c|
\equiv \Phi,
\qquad
N_2^c=N_3^c=N_V^c=0.
\label{eq:inflaton-direction}
\end{equation}
Thus, $N_1^c$ provides the ordinary-family sneutrino component of the
inflationary direction, whereas $\overline N_V^c$ supplies the conjugate
field required to satisfy $D$-flatness. After the waterfall transition,
we assume that the vector-like sector becomes superheavy and effectively
decouples, leaving the light post-inflationary state predominantly aligned
with the ordinary right-handed sneutrino $\widetilde N_1^c$.

Along this trajectory, the inflaton-dependent waterfall mass is controlled by
the coupling $\alpha_1$. The $R$-symmetry-breaking Majorana operator induces
an additional contribution proportional to $\lambda_{11}$, whose magnitude
relative to the contribution generated by $\alpha_1$ is assumed to be small,
\begin{equation}
\left|
\frac{\lambda_{11}}{\alpha_1}
\right|
\frac{M_s}{m_P}
\ll1,
\label{eq:lambda11}
\end{equation}
so that this correction may be neglected in determining the waterfall mass
and the critical inflaton value.

Neglecting the subdominant Majorana contribution, the critical inflaton
value is given by
\begin{equation}
\phi_c
=
2\left[
\frac{\kappa_{\rm w}-1}{|\alpha_1|^2}
\frac{\mu_s^2}{m_P^2}
\right]^{1/4}M.
\label{eq:phi-critical-N1}
\end{equation}
Imposing $\phi_c=M$ determines the coupling $\alpha_1$ as
\begin{equation}
|\alpha_1|
=
4\sqrt{\kappa_{\rm w}-1}\,
\frac{\mu_s}{m_P}.
\label{eq:alpha1-critical}
\end{equation}
Consequently, the hierarchy in Eq.~(\ref{eq:lambda11}) translates into
\begin{equation}
|\lambda_{11}|
\ll
4\sqrt{\kappa_{\rm w}-1}\,
\frac{\mu_s}{M_s}.
\label{eq:lamb2}
\end{equation}

After GUT symmetry breaking, the same $R$-symmetry-breaking operator
generates the Majorana mass of the lightest right-handed neutrino,
\begin{equation}
M_1
=
\lambda_{11}\frac{M^2}{m_P}.
\label{eq:M1}
\end{equation}
Provided that its mixing with the superheavy vector-like sector is
sufficiently suppressed, the light post-inflationary sneutrino state is
predominantly $\widetilde N_1^c$, with
\begin{equation}
m_\phi
\simeq M_1
=
\lambda_{11}\frac{M^2}{m_P}.
\label{eq:inflaton_mass}
\end{equation}
Using $M^2=\mu_s M_s$ together with Eq.~(\ref{eq:lamb2}), one obtains
the corresponding upper scale
\begin{equation}
m_\phi
\ll
4\sqrt{\kappa_{\rm w}-1}\,
\frac{\mu_s^2}{m_P}
\equiv
m_\phi^{\rm ref}.
\label{eq:mphi-bound}
\end{equation}
Thus, $m_\phi^{\rm ref}$ 
characterizes the upper reference scale allowed by the hierarchy in Eq.~(\ref{eq:lamb2}), rather than the physical inflaton mass itself.

\begin{figure}[t]
\centering
\includegraphics[width=\columnwidth,height=\textheight,keepaspectratio]{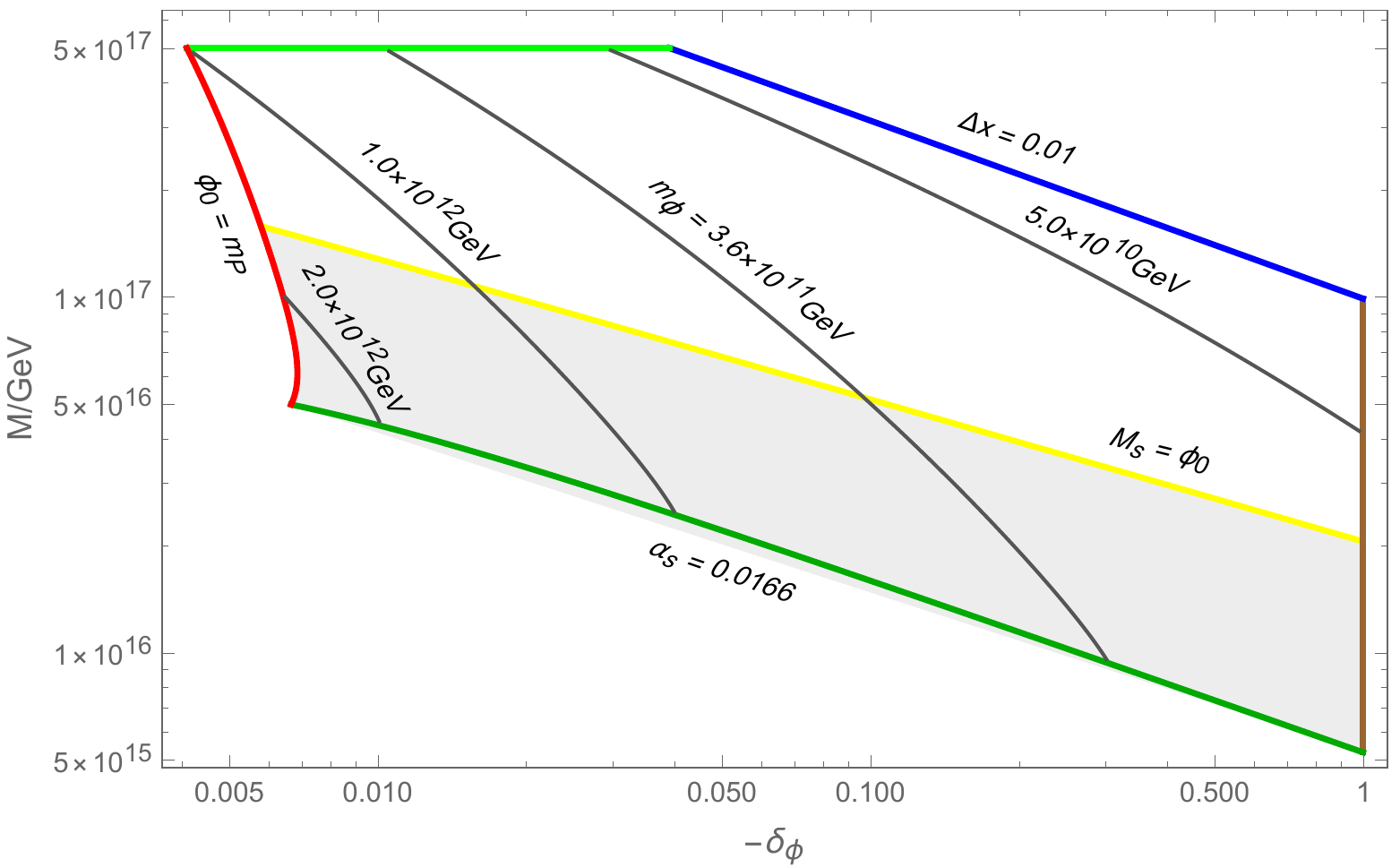}
\caption{
Same as Fig.~\ref{fig:kphi}, but showing contours of constant inflaton mass $m_\phi$, 
and displaying its variation, $3 \times 10^{8} \text{ GeV} \lesssim m_\phi \lesssim 3 \times 10^{12} \text{ GeV}$, within the boundary region.
}
\label{fig:mphi}
\end{figure}

For illustration, we take $\kappa_{\rm w}=2$ and adopt
\begin{equation}
m_\phi = \frac{m_\phi^{\rm ref}}{100} 
\end{equation}
as a representative benchmark satisfying the hierarchy in
Eq.~(\ref{eq:lambda11}). This yields
\begin{equation}
3\times10^{8}~{\rm GeV}
\lesssim m_\phi
\lesssim
3\times10^{12}~{\rm GeV}
\end{equation}
over the phenomenologically allowed parameter space.
Figure~\ref{fig:mphi} displays contours of constant physical inflaton
mass $m_\phi$ for this benchmark choice.

\subsection{Inflaton Decay and Reheating}
The inflaton reheats the Universe through the Yukawa interactions
contained in the renormalizable superpotential
\begin{equation}
W_Y = y_{aj}^{(u,\nu)}\,10_a\,\overline5_j\,\overline5_h .
\label{eq:WY}
\end{equation}
Since the post-inflationary inflaton is predominantly the scalar component
of the lightest ordinary right-handed neutrino superfield $N_1^c$, its
decays are governed by
\begin{equation}
W_Y
\supset
y^\nu_{1j}N_1^cL_jH_u.
\end{equation}
The corresponding scalar inflaton decays through
\begin{equation}
\phi\rightarrow
L_j+\widetilde H_u,
\qquad
\phi\rightarrow
\widetilde L_j+H_u.
\end{equation}
Neglecting final-state masses, the total tree-level decay width is
\begin{equation}
\Gamma_\phi
=
\frac{m_\phi}{4\pi}
\sum_j|y^\nu_{1j}|^2
\equiv
\frac{y_{\rm eff}^2}{4\pi}m_\phi,
\label{eq:GammaPhi}
\end{equation}
where
\begin{equation}
y_{\rm eff}^2
\equiv
\sum_j|y^\nu_{1j}|^2.
\end{equation}

The decay of the inflaton reheats the Universe once
$H\simeq\Gamma_\phi$. Assuming instantaneous thermalization, the
reheating temperature is given by
\begin{equation}
T_r
\approx
\left(\frac{90}{\pi^2g_*}\right)^{1/4}
\sqrt{\Gamma_\phi m_P},
\label{eq:TR}
\end{equation}
where $g_*=228.75$ for the MSSM. Substituting
Eq.~(\ref{eq:GammaPhi}) yields
\begin{equation}
T_r
\simeq
0.45\,
\sqrt{\frac{y_{\rm eff}^2}{4\pi}\,
m_\phi m_P}.
\label{eq:TR2}
\end{equation}
Equivalently, the effective Yukawa coupling required for a given
reheating temperature is
\begin{equation}
y_{\rm eff}
\simeq
\frac{\sqrt{4\pi}}{0.45}\,
\frac{T_r}{\sqrt{m_\phi m_P}}.
\label{eq:yeff-Tr}
\end{equation}
Thus, for fixed $T_r$, the required inflaton Yukawa coupling scales as
$y_{\rm eff}\propto m_\phi^{-1/2}$. For the choice
$T_r=10^6~{\rm GeV}$ adopted throughout our numerical analysis and the
benchmark $m_\phi=m_\phi^{\rm ref}/100$, 
the required effective Yukawa coupling lies approximately in the range
\begin{equation}
3\times10^{-9}
\lesssim
y_{\rm eff}
\lesssim
3\times10^{-7},
\label{eq:yeff-range}
\end{equation}
over the phenomenologically viable parameter space.

Although the couplings $y^\nu_{1j}$ form
part of the ordinary neutrino Yukawa matrix and therefore enter the type-I
seesaw relation, they need not provide the dominant
contribution to the observed light-neutrino masses. Defining the effective
neutrino mass parameter
\begin{equation}
\widetilde m_1
\equiv
\frac{v_u^2}{M_1}
\sum_j|y^\nu_{1j}|^2
=
\frac{v_u^2}{M_1}y_{\rm eff}^2,
\label{eq:mtilde1}
\end{equation}
a small $y_{\rm eff}$ corresponds to a weakly coupled lightest
right-handed neutrino whose contribution to the light-neutrino mass matrix
is negligible. The observed neutrino mass splittings may then be generated
predominantly by $N_2^c$ and $N_3^c$. This allows the small decay coupling
required for a low reheating temperature without conflicting with the
type-I seesaw mechanism.

\subsection{Non-thermal Leptogenesis}

Since the inflaton is identified with the lightest right-handed sneutrino
$\widetilde N_1$, its out-of-equilibrium decays directly generate a lepton
asymmetry. The interference between the tree-level decay amplitude and
one-loop diagrams involving the heavier right-handed neutrinos $N_{2,3}$
produces the CP asymmetry $\varepsilon_1$.
For a hierarchical right-handed-neutrino spectrum, the CP asymmetry is
bounded by \cite{Hamaguchi:2001gw,Davidson:2002qv}
\begin{equation}
|\varepsilon_1|
\lesssim
\frac{3}{8\pi}
\frac{m_\phi m_\nu^{\rm max}}{v_u^2},
\label{eq:DI}
\end{equation}
where $m_{\nu}^{\rm max} = 0.05$~eV is the heaviest light-neutrino mass eigenvalue and $v_u \simeq 174$~GeV in the large $\tan \beta$ limit.

Assuming non-thermal production from inflaton decay, the resulting
lepton asymmetry is
\begin{equation}
Y_L
=
\frac{3}{4}
\frac{T_r}{m_\phi}
\,\varepsilon_1,
\label{eq:YL}
\end{equation}
which is partially converted into the observed baryon asymmetry by
electroweak sphaleron processes,
\begin{equation}
Y_B
=
-\frac{8}{23}\,
Y_L.
\label{eq:YB}
\end{equation}
The observed value \cite{Planck:2018vyg,ParticleDataGroup:2024cfk},
\begin{equation}
Y_B^{\rm obs}
\simeq
8.7\times10^{-11},
\end{equation}
therefore constrains the allowed combinations of
$y_{\rm eff}$ and $m_\phi$.
Combining the sphaleron conversion relation with the
Davidson--Ibarra bound yields a lower limit on the reheating
temperature ,
\begin{equation}
T_r \gtrsim
\frac{92\pi}{9}
\frac{v_u^2}{m_\nu^{\rm max}}
Y_B^{\rm obs}
\simeq 10^6~\mathrm{GeV}.
\end{equation}
Motivated both by the lower bound required for successful non-thermal
leptogenesis and by the gravitino constraint in supersymmetric
cosmology, we adopt
$T_r = 10^6~\mathrm{GeV}$
throughout our numerical analysis.
Finally, the condition
\begin{equation}
T_r\ll M_1\simeq m_\phi
\end{equation}
ensures that thermal production of the lightest right-handed
(s)neutrino after reheating is strongly suppressed. The generated lepton
asymmetry therefore originates predominantly from the out-of-equilibrium
decay of the sneutrino inflaton condensate, while thermal washout effects
remain negligible.

\section{Summary and Conclusions}
\label{sec:conclusions}
We have constructed a realization of sneutrino tribrid inflation within
the $R$-symmetric flipped $SU(5)$ grand unified theory. To accommodate a
gauge non-singlet inflaton while satisfying the $D$-flatness condition, we
extended the matter sector with a vector-like pair of multiplets,
$10_V+\overline{10}_V$, and identified the inflaton with a $D$-flat
combination of the right-handed sneutrino direction and the conjugate
vector-like field. An additional $Z_2$ symmetry forbids a direct inflaton
mass term and the renormalizable hybrid-inflation coupling
$S10_H\overline{10}_H$, so that the leading inflationary interaction is the
non-renormalizable operator $S(10_H\overline{10}_H)^2/M_s^2$. The resulting
setup belongs to the class of K\"ahler-driven tribrid inflation models, in
which the slope of the potential and the waterfall instability that
terminates inflation both originate from higher-dimensional supergravity
corrections rather than from radiative effects.

A central feature of our analysis is the distinction between the
superpotential cutoff scale $M_s$ and the reduced Planck mass $m_P$
governing the K\"ahler expansion, in contrast to previous studies of
K\"ahler-driven tribrid inflation that identify the two scales. Allowing
$M_s$ to vary independently substantially enlarges the viable parameter
space and provides a more general and realistic setting for gauge
non-singlet sneutrino inflation. We derived the effective single-field
inflaton potential up to quartic order in $\phi/m_P$, showed that only the
sign combination $\kappa_\phi>0$, $\delta_\phi<0$ yields a phenomenologically
viable, sub-Planckian, red-tilted inflationary trajectory, and verified that
radiative (Coleman--Weinberg) and soft supersymmetry-breaking corrections
remain negligible throughout the parameter space of interest.

Confronting the model with the ACT DR6/Planck determination
$n_s=0.9734\pm0.0034$, together with the $2\sigma$ bound on the running of
the spectral index, $\alpha_s\leq0.0166$, and the theoretical requirements
of sub-Planckian field values and perturbative control ($\phi_0<m_P$,
$\phi_0<M_s$), we identified broad regions of viable parameter space. Within
this region, the symmetry-breaking scale typically satisfies
$M\gtrsim\text{few}\times10^{16}$~GeV, slightly above the conventional MSSM
unification scale, while the cutoff scale spans
$0.003\lesssim M_s/m_P\lesssim1$ and the inflationary energy scale ranges
over $10^{14}\lesssim\mu_s/{\rm GeV}\lesssim10^{16}$. The predicted
tensor-to-scalar ratio spans $3\times10^{-10}\lesssim r\lesssim0.02$, with a
significant fraction of the viable parameter space yielding
$r\gtrsim10^{-3}$ -- within reach of forthcoming CMB $B$-mode experiment
such as LiteBIRD -- while the predicted running remains positive
and consistent with the current ACT bound throughout.

We further examined the post-inflationary dynamics, identifying the
inflaton with the lightest ordinary right-handed sneutrino embedded in a
conventional type-I seesaw sector, with the vector-like neutrino states
decoupling as a superheavy Dirac pair. Reheating proceeds through the
inflaton's Yukawa coupling to the lepton and up-type Higgs superfields, and
for the representative reheating temperature $T_r=10^6$~GeV -- motivated
jointly by the lower bound from successful non-thermal leptogenesis and by
the gravitino constraint -- we obtained a physical inflaton mass in the
range $3\times10^{8}\lesssim m_\phi\lesssim3\times10^{12}$~GeV and a
correspondingly small effective Yukawa coupling,
$3\times10^{-9}\lesssim y_{\rm eff}\lesssim3\times10^{-7}$. Because this
coupling need not dominate the light-neutrino mass matrix, the observed
neutrino mass splittings can instead be generated predominantly by the
heavier right-handed neutrinos, without conflicting with the seesaw
mechanism. The out-of-equilibrium decay of the sneutrino inflaton condensate
generates a lepton asymmetry that is subsequently reprocessed by
electroweak sphalerons into the observed baryon asymmetry, with negligible
thermal washout given $T_r\ll M_1$.

Altogether, the flipped $SU(5)$ tribrid inflation model presented here
provides a unified and economical framework that simultaneously accounts
for GUT-scale symmetry breaking, the seesaw origin of neutrino masses, a
CMB-consistent inflationary epoch, and the observed baryon asymmetry via
non-thermal leptogenesis, while offering a potentially observable
primordial gravitational-wave signal for upcoming CMB polarization
experiments.

\section*{Acknowledgments}
The authors thank Maria Mehmood for her valuable contributions during the initial stages of this work. M.U.R. thanks Stefan Antusch for useful discussions.  S.O.A. and M.U.R. extend their appreciation to the Deanship of
Scientific Research, Islamic University of Madinah, Saudi
Arabia, for funding this research work.
\newpage

\bibliographystyle{apsrev4-1}
\bibliography{bibliography}

\end{document}